\RequirePackage{fix-cm}
\documentclass{article}    

\usepackage{graphicx} 
\usepackage{amsfonts,amsmath,amssymb} 
\usepackage{multirow} 
\usepackage{hyperref,graphicx} 
\usepackage{setspace} 
\usepackage{float}

\begin{document} 

\title{An Exploration of Spatial Radiomic Features in Pulmonary Sarcoidosis}

\author{Sarah M. Ryan \and 
		Tasha Fingerlin, \and 
		Nabeel Hamzeh \and 
		Lisa Maier \and 
		Nichole Carlson}

%\institute{Sarah M. Ryan, MS
%		\at Department of Biostatistics and Informatics, University of Colorado Denver\\
%			Tel.: 319-929-9322
%			\email{sarah.m.ryan@ucdenver.edu}
%		\and Tasha Fingerlin, PhD
%		\at National Jewish Health
%		\and Nabeel Hamzeh, MBBS
%		\at University of Iowa
%		\and Lisa Maier, MD, MSPH, FCCP
%		\at National Jewish Health
%		\and Nichole Carlson, PhD
%		\at Department of Biostatistics and Informatics, University of Colorado Denver

\maketitle 

\begin{abstract} 
Sarcoidosis is a rare, multi-systemic, inflammatory disease, primarily affecting the lungs. High-resolution computed tomography (CT) scans are used to clinically characterize pulmonary sarcoidosis. In the medical imaging field, there is growing recognition to switch from visual examination of CT images to more rapid, objective assessments of the abnormalities. In this work, we explore the usefulness of various objective measures of spatial heterogeneity---fractal dimension, Moran's $I$, and Geary's $C$ ---for distinguishing between abnormal sarcoidosis and normal lung parenchyma. 

CT data for N=58 sarcoidosis subjects enrolled at National Jewish Health were obtained from the GRADS study. CT data for N=101 control patients were obtained from the COPDGene study. Radiomic measures were computed for each two-dimensional slice of a given scan, in the axial, coronal, and sagittal planes. Functional regression was applied to identify lung regions where CT nodules tend to proliferate. 

Moran's $\mathcal{I}$, Geary's C and fractal dimension significantly differentiate between subjects with and without sarcoidosis throughout the majority of the lung, with disease abnormalities most apparent in the top axial, middle coronal, and outer sagittal regions. A trend appeared across Scadding stages, with CT scans from patients with Scadding stages I and III appearing the healthiest, and Scadding stage IV appearing the least healthy. 

The radiomic measures and techniques presented herein successfully characterize CT images in sarcoidosis by objectively and efficiently approximating what we know about the pathology of sarcoidosis.  

%\keywords{Computed Tomography \and Pulmonary Sarcoidosis \and Fractal Dimension \and Moran's {\em I} \and Geary's {\em C} \and Mat\'{e}rn parameters}
\end{abstract} 

\newpage
\section{Introduction} 
\label{intro}
Sarcoidosis is a rare, multi-systemic, inflammatory disease, primarily affecting the lungs, lymph nodes, eyes and skin. Although the etiology of sarcoidosis is unknown, the combination of environmental and genetic factors are important in understanding disease risk. Sarcoidosis affects 4.7 to 64 per 100,000 persons, with seasonal peaks in the spring, and the highest rates found in northern European and African-American females (Valeyre et al. 2014). While spontaneous remission may occur in up to two thirds of subjects in certain populations (Nunes et al. 2005), mortality rates due to respiratory failure are rising in the US (Swigris et al. 2011). 

Pulmonary sarcoidosis is a common presentation of sarcoidosis, as lung involvement affects 95\% of those with the disease. Symptoms include cough, chest discomfort, and poor pulmonary function as measured by FVC, FEV1, and total lung capacity (Butler and Keane 2016). Clinical and radiological presentations are used for diagnosis of pulmonary sarcoidosis, with chest radiographs showing abnormalities in 90\% of cases (Valeyre et al. 2014). The Scadding system of staging classifies subjects with sarcoidosis into five stages: stage 0 (no radiograph abnormalities), stage 1 (bilateral hilar lymphadenopathy (BHL) alone), stage 2 (pulmonary infiltration with BHL), stage 3 (pulmonary infiltration without BHL), and stage 4 (pulmonary fibrosis with volume loss). Pulmonary infiltration is characterized by small nodular opacities that track along the bronchovascular bundle; these opacities are usually bilateral, symmetrical, and predominately in the central regions and upper lobes. BHL is an enlargement of the lymph nodes that are in or surrounding the mediastinum. A wide range of additional chest radiographic patterns may be observed, including: focal alveolar and ground-glass opacities, pleural thickening or calcification, upper lobe volume loss with hilar retraction, coarse linear bands, and bullae. In general, as the disease worsens, these patterns become more prominent (Nunes et al. 2005). However, despite the nomenclature, there is no sequential ordering of Scadding stages, and subjects may or may not experience any of the stages at various times during their diagnosis.

High-resolution computed tomography (CT) scans are used to clinically characterize sarcoidosis. Some studies have shown the ability of CT scans to differentiate between patients with interstitial lung disease and healthy controls (Sluimer et al. 2006). However, these studies rely on visual examination of CT scans by pulmonologists and radiologists. Visual examination of disease on medical images is time consuming and known to have poor inter-rater reliability (Wormanns et al. 2000). Thus, in the field of medical imaging, there is a growing recognition that switching to automated reads of CT scans would provide a number of advantages, including a more rapid and objective assessment of the abnormalities.

Unlike sarcoidosis, objective assessments are increasingly finding application for quantification of disease severity on CT scans of subjects with emphysema. Measures and approaches include percentage of low attenuation, mean lung density, histogram analysis, and Adaptive Multiple Feature Method (Uppaluri et al. 1997). However, there is a lack of objective methods to quantify disease severity of sarcoidosis, due to disease relative rarity and unique manifestations on CT scans. That is, sarcoidosis presents on CT scans as areas of high attenuation (opposite of emphysema), which, problematically, is similar to the presentation of airways and vasculature of the lung. Often, unwanted tissues can be masked from CT scans in pre-processing; yet, masking of airways and vasculature in subjects with sarcoidosis can be difficult (Nardelli et al. 2015). As such, the presentation of sarcoidosis in the lungs is muddled with `background noise', which limits the applicability of standard objective CT quantification biomarkers, such as percentage of attenuation, to quantify disease severity of sarcoidosis.

Due to the presence of visual patterns in pulmonary CT images of subjects with sarcoidosis, we turn to spatial measures to objectively quantify and describe these data. In this work, we will investigate various putative spatial radiomic biomarkers, including  Moran's $\mathcal{I}$, Geary's C, fractal dimension and Mat\'{e}rn parameters, to differentiate between subjects with and without sarcoidosis. To identify regions of the lung where disease tends to proliferate, a slice-by-slice functional analysis will be performed in the three anatomical planes (axial, coronal, and sagittal). The axial (or transverse) plane divides the body into top and bottom portions (i.e. cranial and caudal); the coronal plane divides the body into back and front (i.e. posterior and anterior); the sagittal plane divides the body into right and left portions. The novelty of our approach lies in comprehensively examining spatial radiomic features in sarcoidosis tissue of the lung, and identifying lung regions of interest in a time- and space-efficient computation.

\section{Spatial Radiomic Measures}
\label{sec:1}

\subsection{ Moran's $\mathcal{I}$ }
Spatial autocorrelation is a measure of similarity between values located in space. One popular attempt at summarizing spatial autocorrelation is  Moran's $\mathcal{I}$.  Moran's $\mathcal{I}$ relies on the product moments of values located in space. However, only the product moments between pixels at a specific distance are considered in the calculation. The collection of distances included is known as the ``neighborhood''. 

 Moran's $\mathcal{I}$ is, arguably, the most common measure of spatial autocorrelation, and is often used in exploratory data analysis for areal data (Banerjee et al. 2014). While there are proposed modifications and alternatives to Moran's $\mathcal{I}$ (Maruyama 2015), we rely on the original estimate as proposed by Moran (1950). 
Moran's $\mathcal{I}$ is given by:
\begin{align}
\mathcal{I}&=\frac{n}{\sum_{i=1}^{n}\sum_{j=1}^{n}{w_{ij}}}\frac{\sum_{i=1}^{n}\sum_{j=1}^{n}{w_{ij}(y_i-\bar{y})(y_j-\bar{y})}}{\sum_{i=1}^{n}{(y_i-\bar{y})^2}}
\end{align}
where $n$ is the number of non-missing pixels; $w_{ij}$ is an element in the spatial weight matrix ($W$); $y_i$ is the HU intensity at a single pixel; $\bar{y}$ is the mean HU intensity over the whole slice, $\tilde{y}=y_i-\bar{y}$; and $I$ is the identity matrix. 

Various spatial weight matrices can be used in this estimation. The original and most common weight matrices for  Moran's $\mathcal{I}$ are the four or eight nearest neighbor matrices. The range of  Moran's $\mathcal{I}$ depends on its spatial weight matrix. As shown by Maruyama (2015),  Moran's $\mathcal{I}$ is bound between the smallest and second-largest eigenvalues of its spatial weight matrix. Regardless of the weight matrix used, values around zero are indicative of no spatial autocorrelation, with higher values indicating a positive spatial autocorrelation. Since CT nodules begin to conglomerate in sarcoidosis subjects, we hypothesize that sarcoidosis subjects will have higher values for  Moran's $\mathcal{I}$ as compared to healthy controls.

\subsection{Geary's C}
Geary's C (Geary 1954) is also a measure of spatial autocorrelation. It differs from  Moran's $\mathcal{I}$ in that it is more sensitive to smaller scale spatial structures, due to its reliance on squared differences, rather than on product moments.  Geary's C is given by: 
\begin{align}
C&=\frac{(n-1)}{2\sum_{i=1}^{n}\sum_{j=1}^{n}}\frac{\sum_{i=1}^{n}\sum_{j=1}^{n}{w_{ij}{(y_i-y_j)^2}}}{\sum_{i=1}^{n}{(y_i-\bar{y})^2}}
\end{align}
where $L=D-W$; $D$ is the diagonal matrix; $n$ is the number of non-missing pixels; $w_{ij}$ is an element in the spatial weight matrix ($W$); $y_i$ is the HU intensity at a single pixel; $\bar{y}$ is the mean HU intensity over the whole slice; and $\tilde{z}=y_i-y_j$. Similar to Moran's $\mathcal{I}$, various spatial weight matrices can be used. 

Geary's C ranges from 0 to 2, with values close to zero indicating strong positive spatial autocorrelation, values close to two indicating strong negative spatial autocorrelation, and a value of one indicating no spatial autocorrelation. Since Geary's C is inversely proportional to  Moran's $\mathcal{I}$, we hypothesize that sarcoidosis subjects will have lower values of Geary's C as compared to healthy controls.

\subsection{Fractal dimension}
Fractal dimension differs from Moran's $\mathcal{I}$ and Geary's C, in that it is inherently a measure of point-level spatial data. Commonly referred to as a property of statistical self-similarity or complexity, fractal dimension is also a measure of roughness, as the spatial scale becomes infinitesimally small (Gneiting et al. 2010). 

There are various methods to estimate fractal dimension on two-dimensional spatial domains. However, Gneiting et al. (2010) showed that the line transect madogram estimator, based on the  power ($\hat{V}$) and variation estimates ($\hat{D}$) is both efficient and robust. Fractal dimension estimated with the line transect madogram is defined as:
\begin{align}
\text{D} &= 1+\text{median}\{\hat{D}_{k}\}  
\end{align}
where $k$ represents the index for the row or column line transect, and $\hat{D}_k$ is the fractal dimension estimate for a single line transect, defined as: 
\begin{align}
\hat{D}_k &= 2-\frac{\log\hat{V}_k(2)-\log\hat{V}_k(1)}{\log 2}\\
\hat{V}_k(l) &= \frac{1}{2(m_k-l)}\sum_{\forall i,j \in |i-j|=l}{\left| z_{i}-z_{j}\right|}
\end{align}
where $\hat{V}_k(l)$ represents the variation in a single line transect at the amount of lag, $l$ (or difference in distance between two pixels, $i$ and $j$), and $m_k$ is the number of differences in a single line transect $k$ such that $|i-j|=l$. 

In two-dimensional space, such as a slice of a CT scan, fractal dimension ranges between 2 and 3. Low fractal values are indicative of the lack of self-similarity patterns, and may be characterized as `smooth'; whereas high values indicate self-similarity patterns, and may be characterized as `rough'. 
Since sarcoidotic tissue disrupts the inherent self-similarity pattern seen in the healthy lung, we hypothesize that fractal dimension will be lower in subjects with sarcoidosis as compared to healthy controls.

\section{Methods}
\label{sec:methods}
\subsection{Data acquisition}
The CT data for sarcoidosis subjects enrolled at National Jewish Health were obtained from the Genomic Research in Alpha-1 Antitrypsin Deficiency and Sarcoidosis (GRADS) study. The GRADS study is a multi-center, observational cohort study exploring the role of the microbiome and genome in subjects with Alpha-1 Antitrypsin Deficiency and/or Sarcoidosis (Moller et al. 2015). Approximately 358 sarcoidosis subjects were enrolled across the entire program after undergoing informed consent for a protocol that was approved by the sites IRB; however, the present study focuses on subjects with sarcoidosis seen at National Jewish Health (NJH) in Denver, Colorado. At NJH, 79 subjects with varying stages of sarcoidosis were enrolled in the study, including N=58 subjects with Scadding stage I-IV chest radiographs. As part of the GRADS study, uniform clinical data was obtained including self-administered questionnaires, pulmonary function testing, a chest radiograph (for Scadding staging classification), and a research chest CT. All research chest CTs were acquired using Siemens SOMATOM Definition, with the participants in the supine position during breath holding at end inspiration, and the following machine parameters: 500msec exposure time, B35f kernel, and 0.75mm thickness (Moller et al. 2015).

CT data for control subjects were obtained from the COPDGene study. COPDGene is a cross-sectional prospective cohort exploring epidemiological and genetic characteristics in smokers with and without COPD (Regan et al. 2010).  Since smoking is protective for sarcoidosis, few subjects had a history of smoking; hence, we included only only healthy, non-smoking subjects in the present study. These subjects received chest CT scans using the same protocol as the GRADS study. All CT scans and data in this study were provided with a de-identified study ID number without any identifying information.  

\subsection{Data-processing}
CT scans for subjects with sarcoidosis were received in NIfTI-1 file format. Healthy controls were received in DICOM format, then converted to NIfTI-1 file format using function, `dicom2nifti', available in the `oro.dicom' v0.5.0 R package. All scans were unscaled. Masks were applied to segment the right and left lungs, and checked for quality using visual inspection and slice standard deviations. All Hounsfield units (HU) outside the standard CT window setting for pulmonary parenchyma (center: -500 HU, width: 1800 HU) (Teague et al. 2012) were removed due to inaccurate masking. Slices with less than 10,000 pixels (approximately 5\% or less of total slice) were discarded due to small image size and inaccurate estimates. To reduce computation time, every third slice for every CT scan was summarized by fractal dimension,  Moran's $\mathcal{I}$, and Geary's C in the three anatomical planes: axial, coronal, and sagittal. 

 Moran's $\mathcal{I}$ and Geary's C were calculated using the eight nearest neighbor adjacency matrix. Fractal dimension was estimated using the line transect madogram estimator. All three--  Moran's $\mathcal{I}$, Geary's C, and fractal dimension-- were custom-coded in RStudio v1.0.136. 

\subsection{Statistical analysis} 
Descriptive statistics (e.g. mean and standard deviation for continuous variables; frequency tables for categorical variables) were used to summarize demographic and pulmonary function data. To test for an association between disease group and gender, a Chi-square test was used. To test for an association between disease group and race (white/non-white), a Fisher's exact test was used, due to small cell sample sizes. Independent two-sample t-tests assuming unequal variance were used to test for a difference in means for all continuous variables between disease groups. 

To determine whether the putative radiomic biomarkers differed throughout the slices of the lungs between subjects with and without sarcoidosis, functional regression was used. Given the imbalance in covariates, the functional regression was extended to adjust for gender, age, and BMI. This analysis was independently performed for fractal dimension,  Moran's $\mathcal{I}$, Geary's C, and Mat\`{e}rn parameters. Mean estimates and standard errors for each parameter were plotted to understand the patterns of the putative spatial biomarkers throughout the lung. All analyses were performed in Rstudio v1.0.136.

\section{Results}
\label{sec:results}
\subsection{Demographics and clinical characteristics}
In this study, 159 people with CT scans were analyzed, of which 58 had sarcoidosis. In the sarcoidosis group, the majority were white (87.9\%), male (55.2\%), with a mean age of 54.1 years (SD: 9.0 years) and a BMI of 28.8 (SD: 6.2). In the control group, the majority were white (94.1\%), female (68.3\%), with a mean age of 62.1 years (SD: 9.1 years) and a BMI of 28.0 (SD: 4.9). The sarcoidosis group had a significantly higher percentage of males (p=0.006), lower mean age (p$<$0.001), and higher height (p=0.001) and weight (p=0.019) (Table~\ref{tab:demo}). Additionally, the sarcoidosis group had a significantly lower percent predicted post-bronchodilator FEV1 (p$<$0.001) and FVC (p$<$0.001)) (Table~\ref{tab:clin}).

\subsection{Behavior of radiomic features for selected representative subjects}
\begin{figure}[H]
\centering
  \includegraphics[width=1.5in]{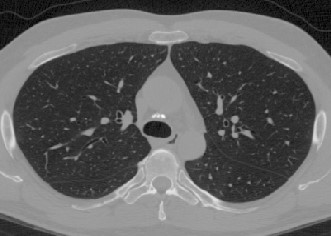}
  \includegraphics[width=1.5in]{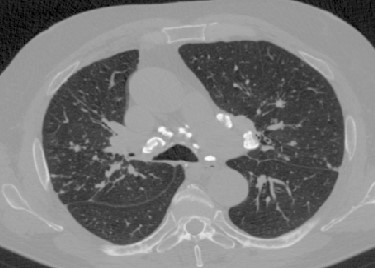}
  \includegraphics[width=1.5in]{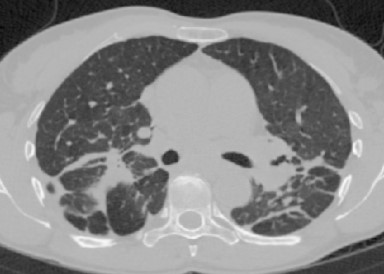}
  \includegraphics[width=1.5in]{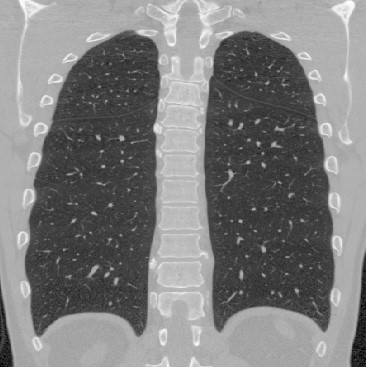}
  \includegraphics[width=1.5in]{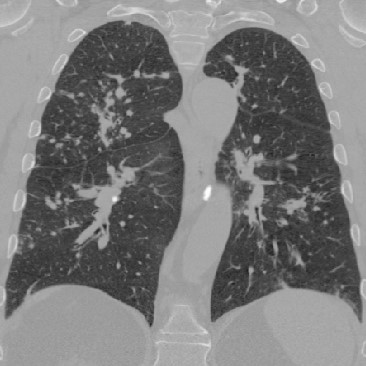}
  \includegraphics[width=1.5in]{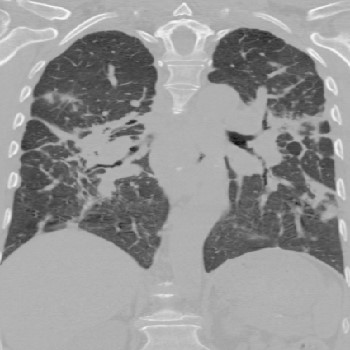}
  \includegraphics[width=1.5in]{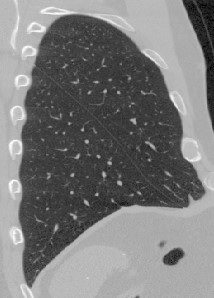}
  \includegraphics[width=1.5in]{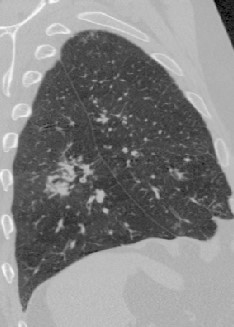}
  \includegraphics[width=1.5in]{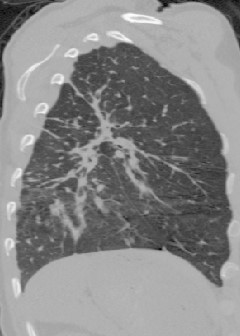}
  \caption[Selected slices of CT scans from representative subjects]{\linespread{1.3}\selectfont{} Slices of CT scans from subjects representative of CT abnormalities, including healthy (column 1), diseased (column 2), and very diseased (column 3). The slices are shown for the 70th percentile axial slice (row 1), 40\% coronal slice (row 2), and 80\% sagittal slice (row 3). }
  \label{fig:views}
\end{figure}

To illustrate the patterns in the radiomic biomarker information obtained across the range of disease, we examine the values from three representative individuals, from healthy, mildly abnormal (Sarc A), and very abnormal (Sarc B) CT scans. As disease becomes more apparent on the CT slices (from healthy to very abnormal) (Figure~\ref{fig:views}), the values for  Moran's $\mathcal{I}$ increase. In contrast, the values for fractal dimension and Geary's C decrease (Tables~\ref{tab:axial}, \ref{tab:coronal}, \ref{tab:sagittal}). An increase in  Moran's $\mathcal{I}$ and a decrease in Geary's C are indicative of more positive spatial autocorrelation; a decrease in fractal dimension $\mathcal{D}$ is indicative of smoother structures. The trend in these parameters are clinically important, since we expect to see more positive spatial autocorrelation and smoother structures in CT slices as disease worsens. 

\subsection{ Moran's $\mathcal{I}$}
\begin{figure}[H]
\centering
  \includegraphics[width=2.25in]{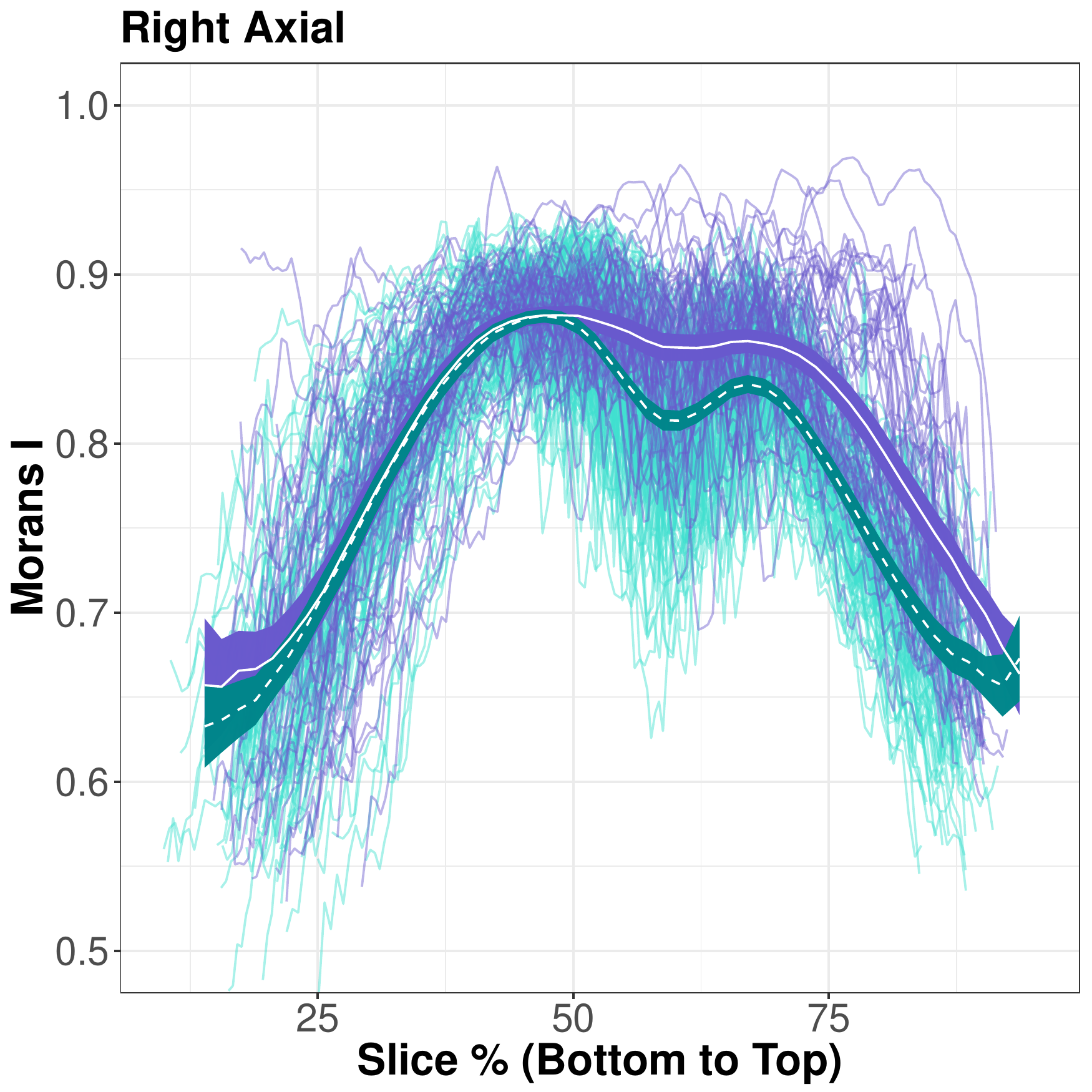}
  \includegraphics[width=2.25in]{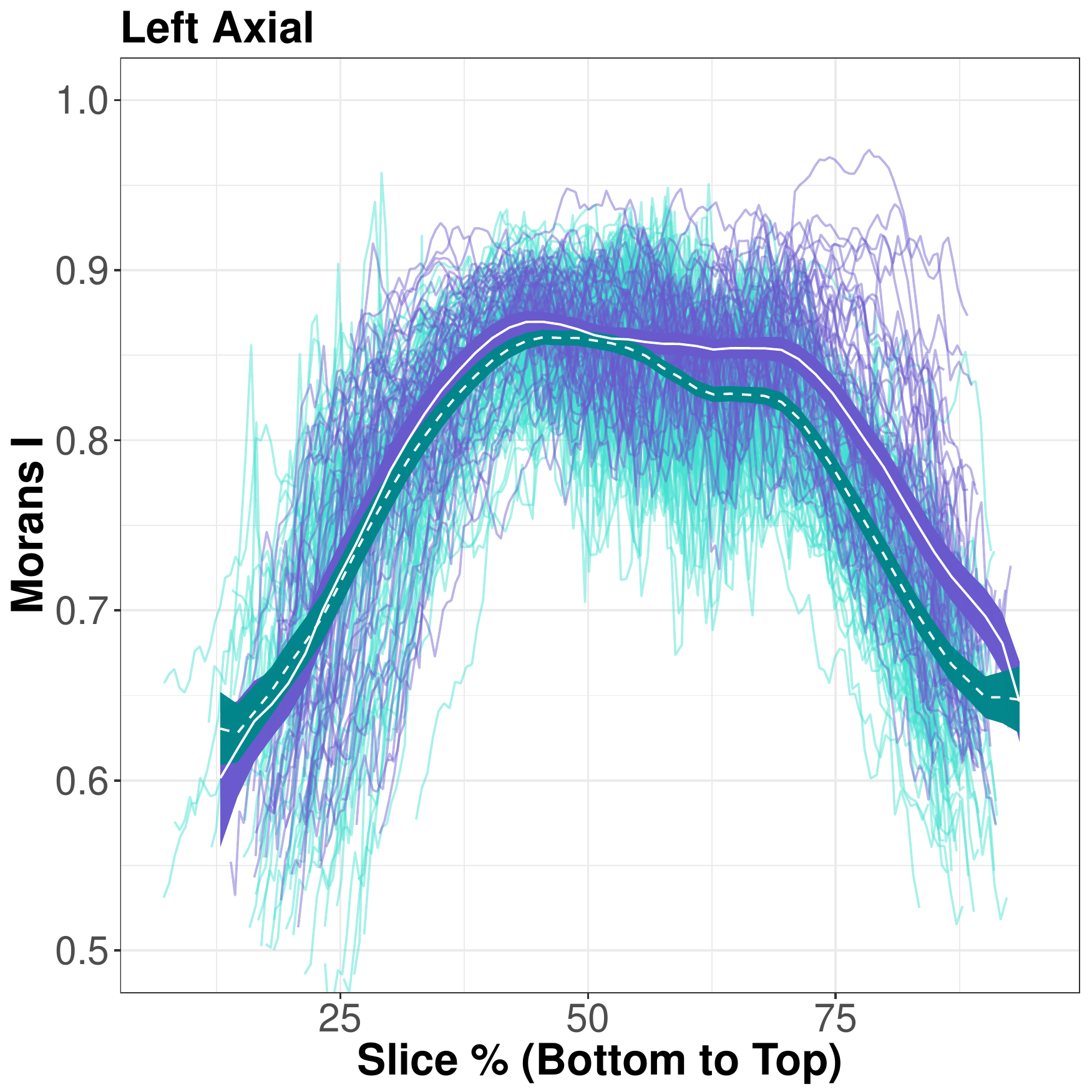}
  \includegraphics[width=2.25in]{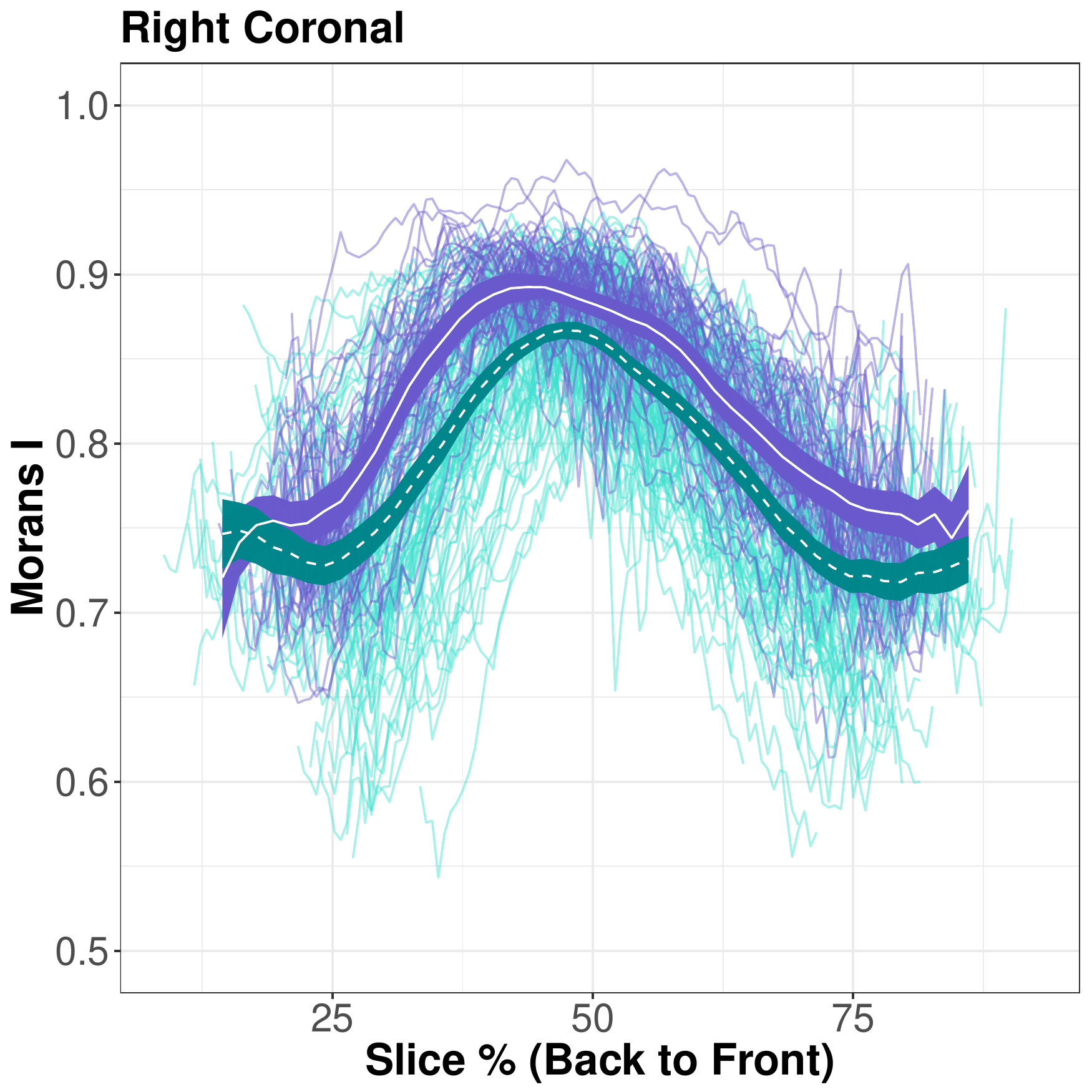}
  \includegraphics[width=2.25in]{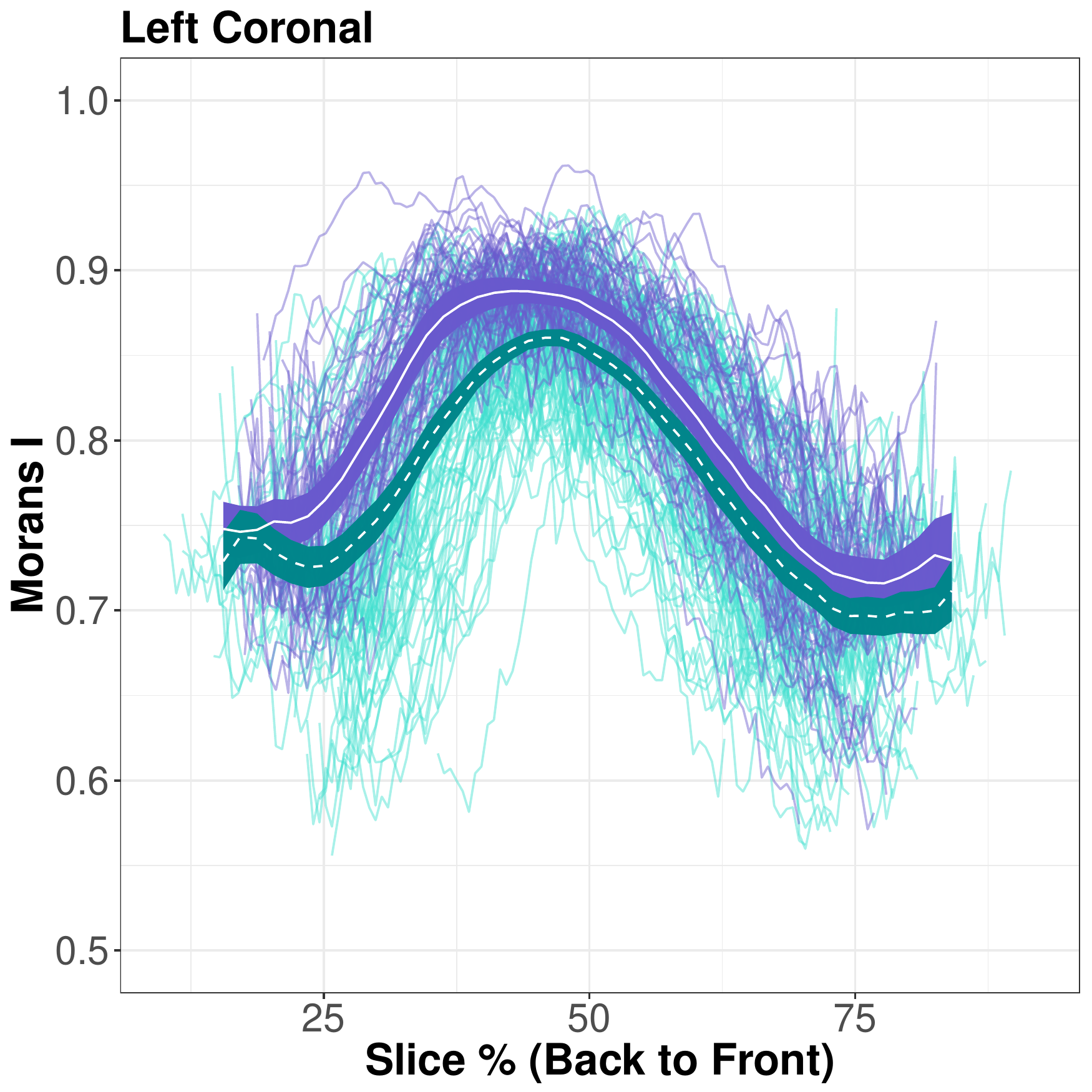}
  \includegraphics[width=2.25in]{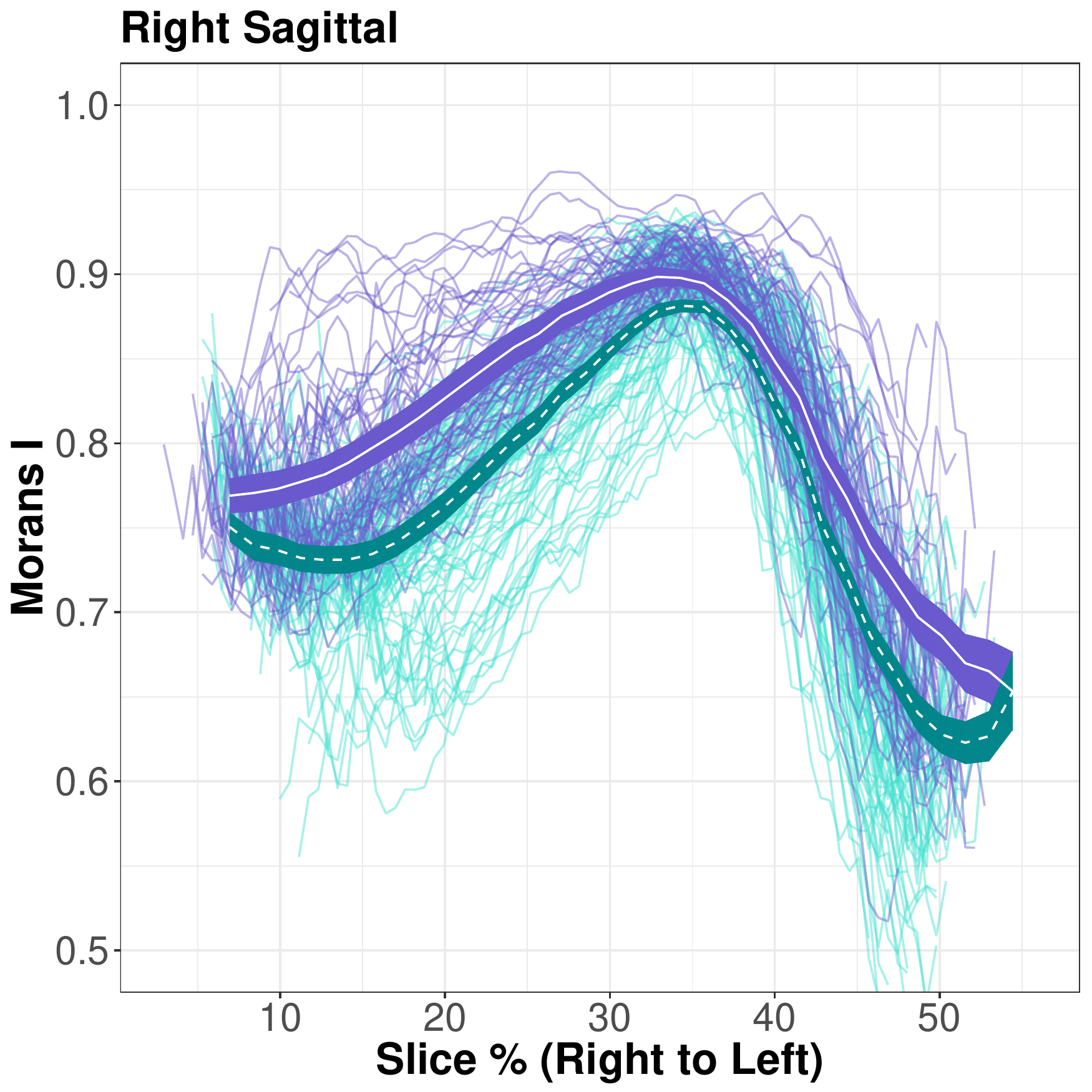}
  \includegraphics[width=2.25in]{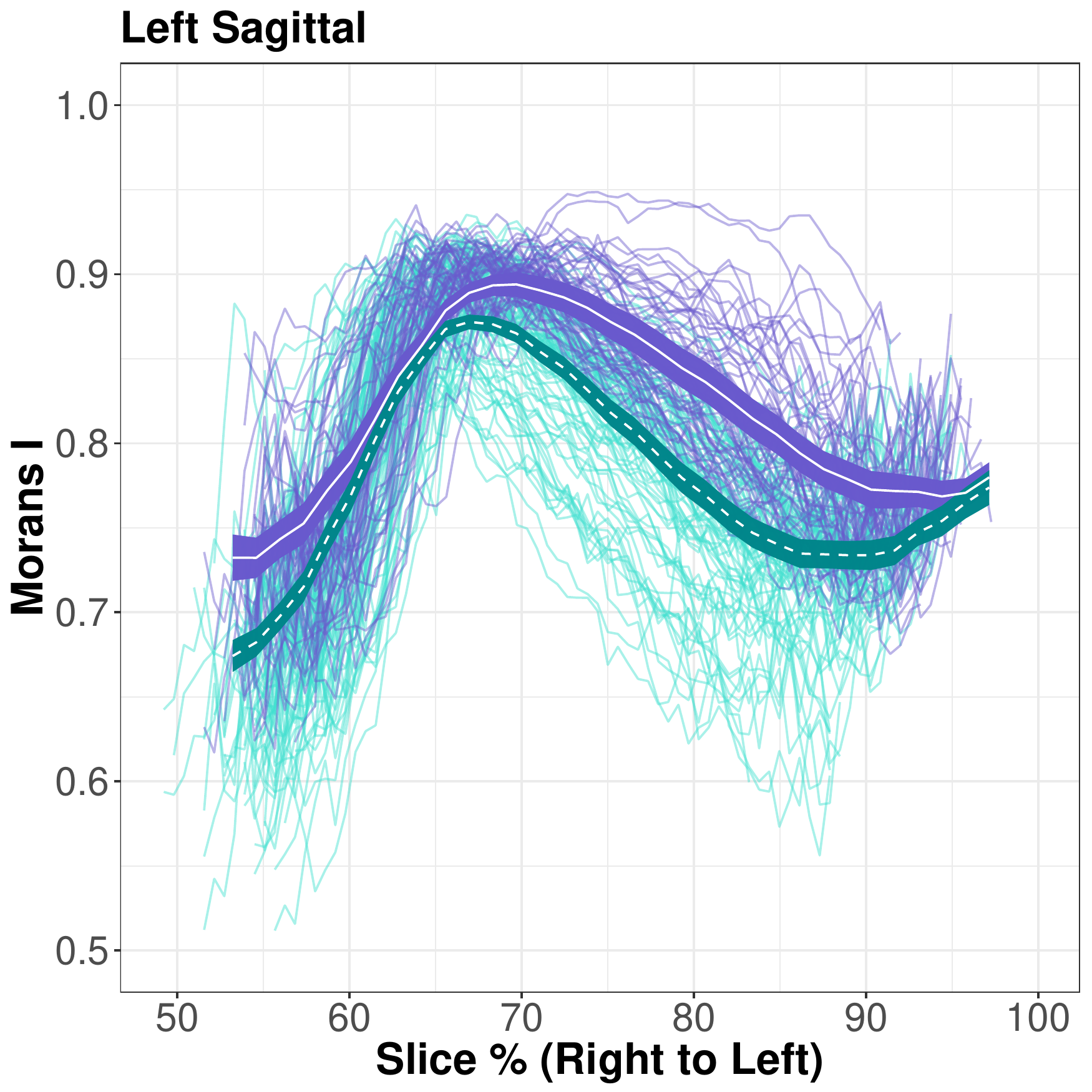}
  \caption[Crude:  Moran's $\mathcal{I}$ throughout the lungs]{\linespread{1.3}\selectfont{}  Moran's $\mathcal{I}$ (with binary eight nearest neighbors weighting) throughout the lung. Bands represent 95\% CIs.  The green dashed lines indicate healthy controls, and the purple solid lines represent subjects with sarcoidosis. }
  \label{fig:mi}
\end{figure}

For all subjects, Moran's $\mathcal{I}$ ranges from 0.45 to 0.97 units, indicative of positive spatial structure. In the axial plane,  Moran's $\mathcal{I}$ is highest in the middle of the lung, and lowest at the bottom and top of the lung. In the coronal plane,  Moran's $\mathcal{I}$ is highest in the middle of the lung, and lowest at the back and front of the lung. In the sagittal plane,  Moran's $\mathcal{I}$ is mid-range on the outside right and left of the lungs, then drops to its lowest at the center of the lungs (Figure~\ref{fig:mi}). These results indicate that the lung shows the most global spatial structure in the middle axial, coronal, and sagittal regions, and the least spatial structure in the bottom/top axial, back/front coronal, and inner sagittal. 

Without adjusting for covariates (Figure~\ref{fig:mi}),  Moran's $\mathcal{I}$ significantly differentiates between subjects with and without sarcoidosis in certain regions of the lung. The areas with the most differentiation between diseased and control subjects are the top axial, middle coronal, and outer sagittal regions. In general, subjects with sarcoidosis have higher  Moran's $\mathcal{I}$ than healthy controls, indicative of more positive spatial structures. These results are similar in both the right and left lungs.

\begin{figure}[H]
\centering
  \includegraphics[width=2.25in]{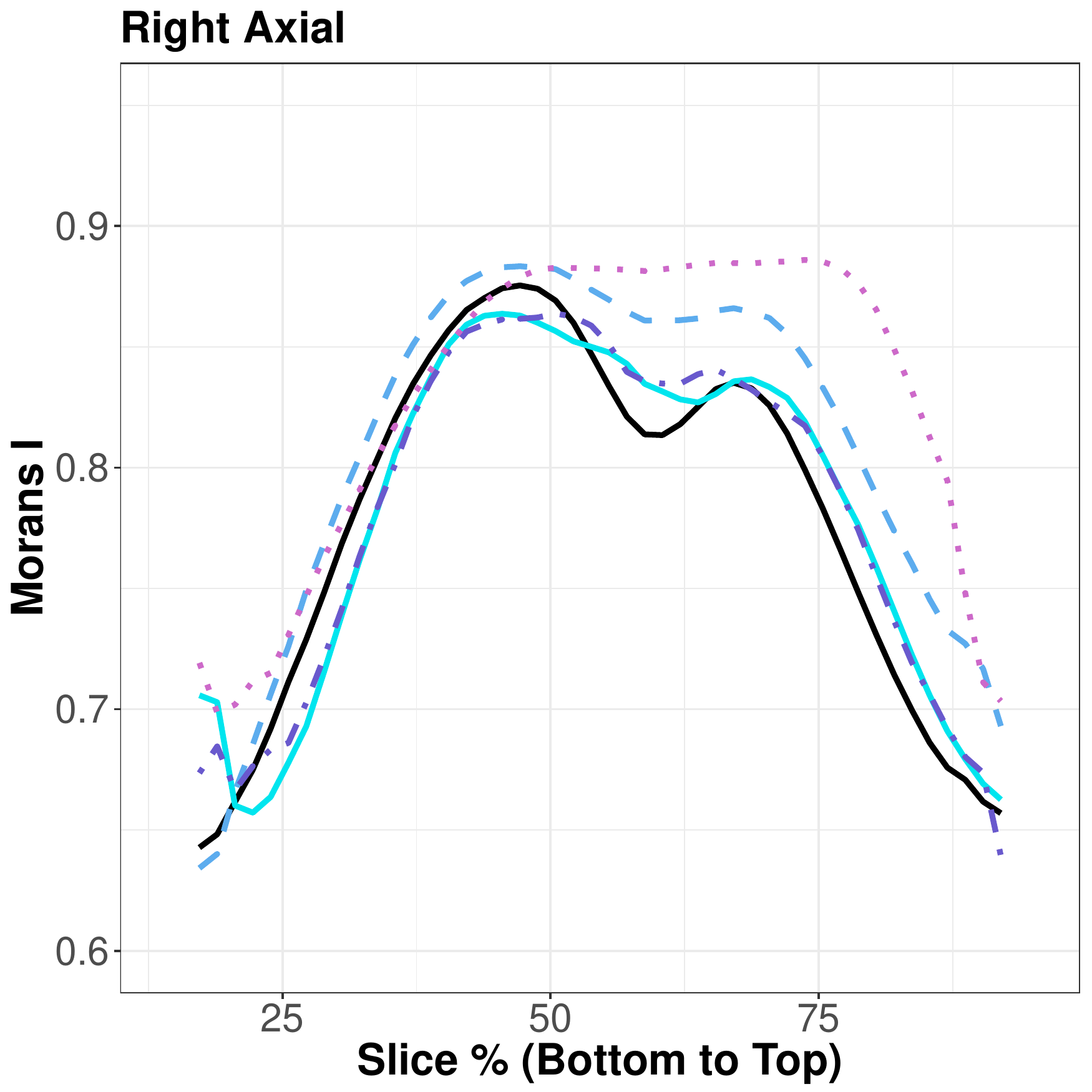}
  \includegraphics[width=2.25in]{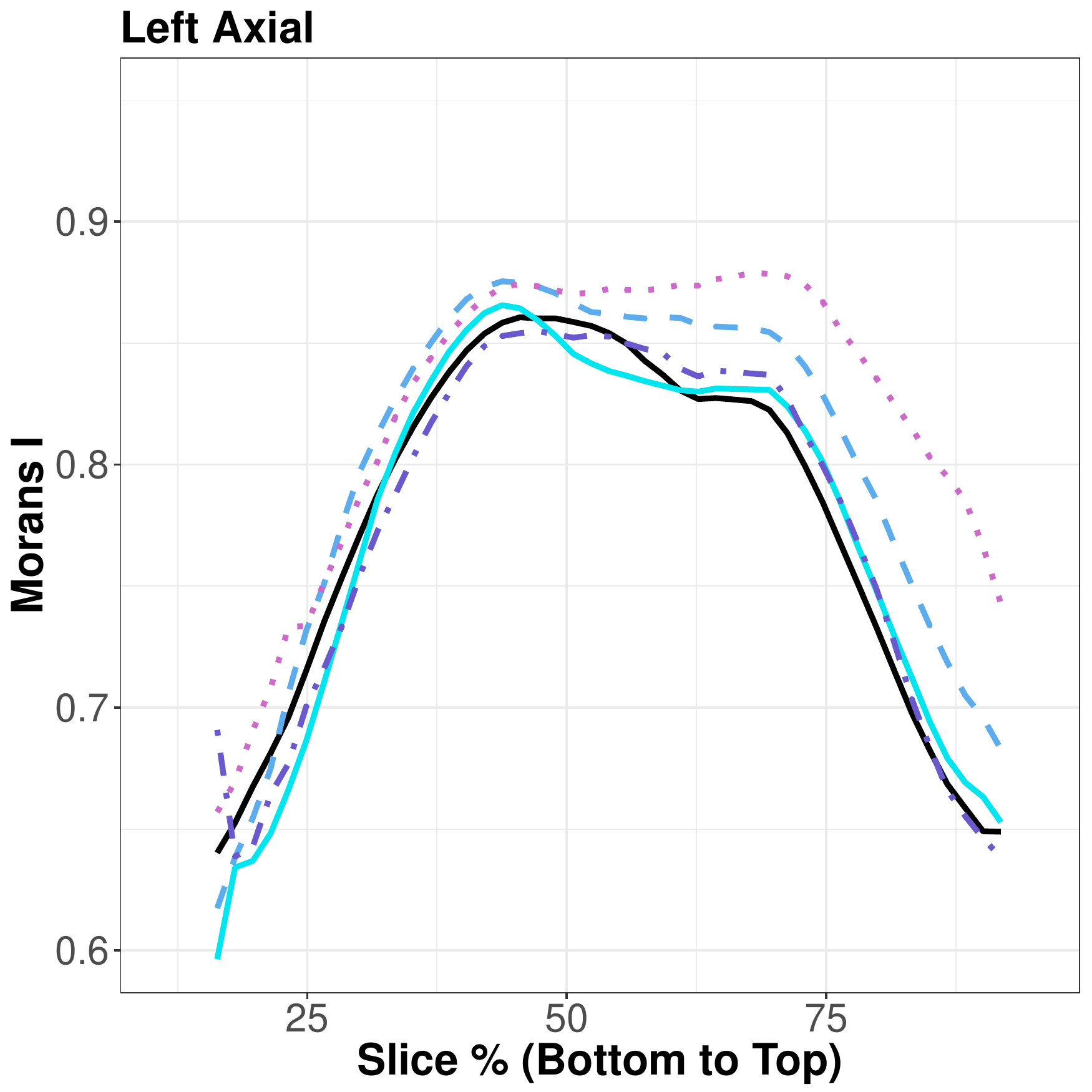}
  \includegraphics[width=2.25in]{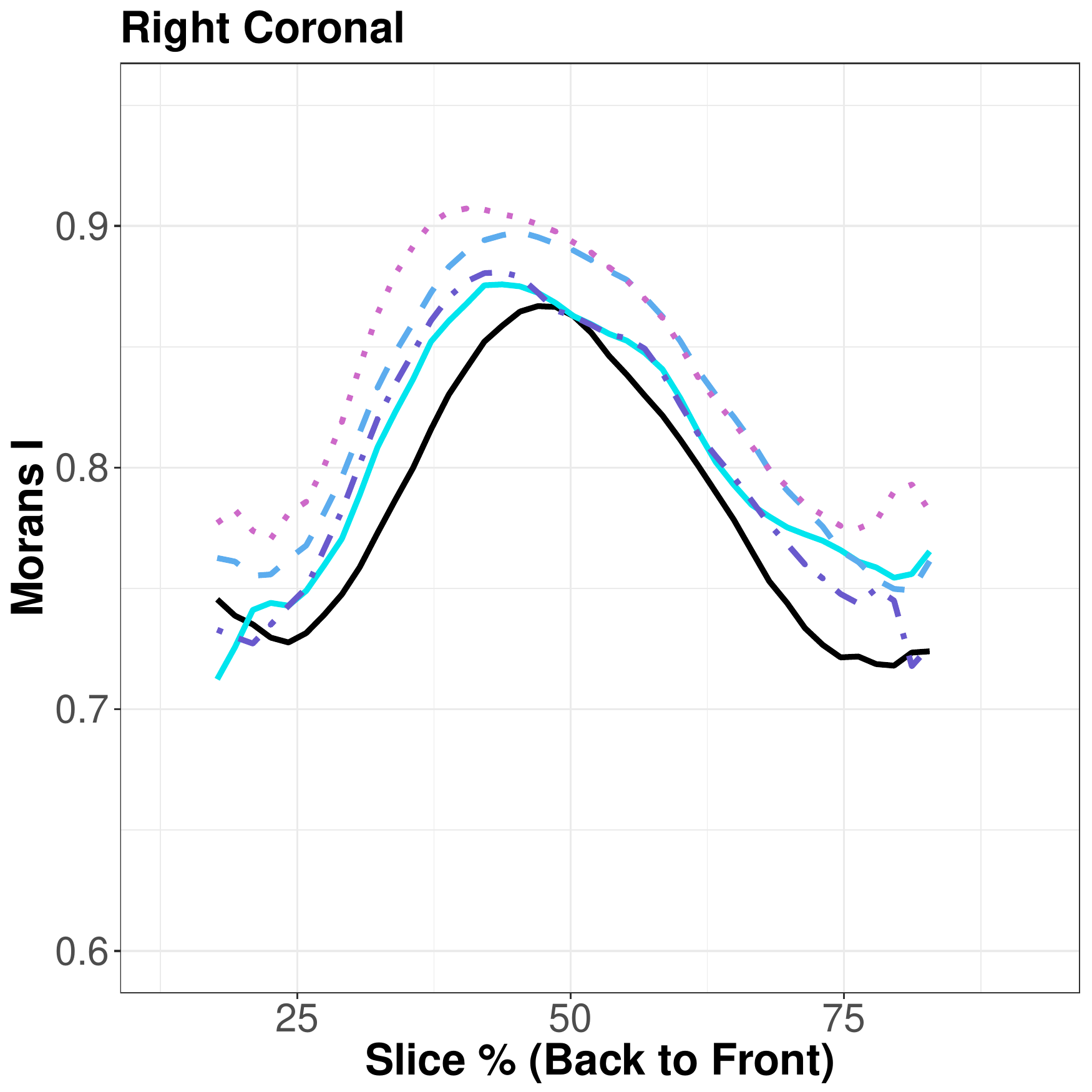}
  \includegraphics[width=2.25in]{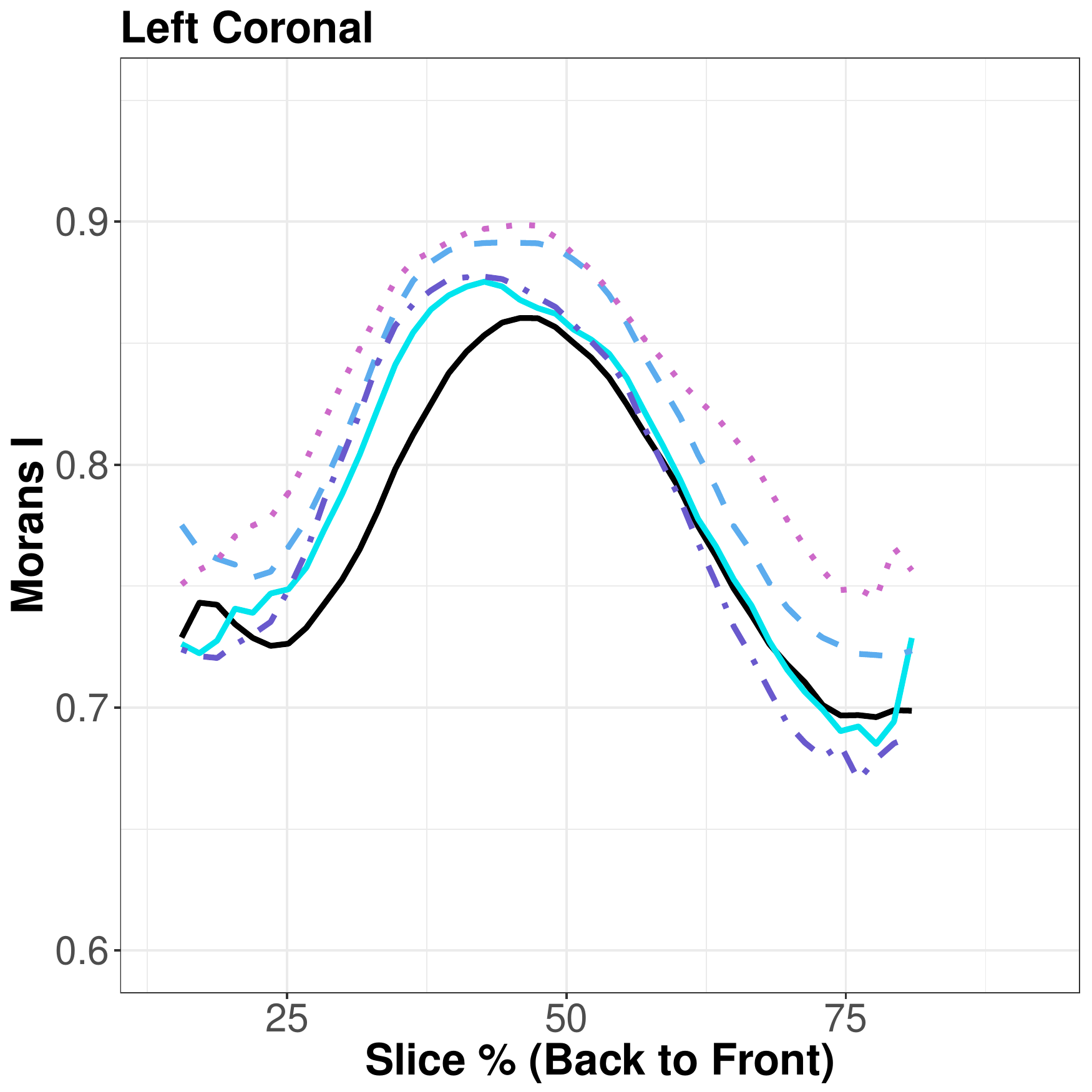}
  \includegraphics[width=2.25in]{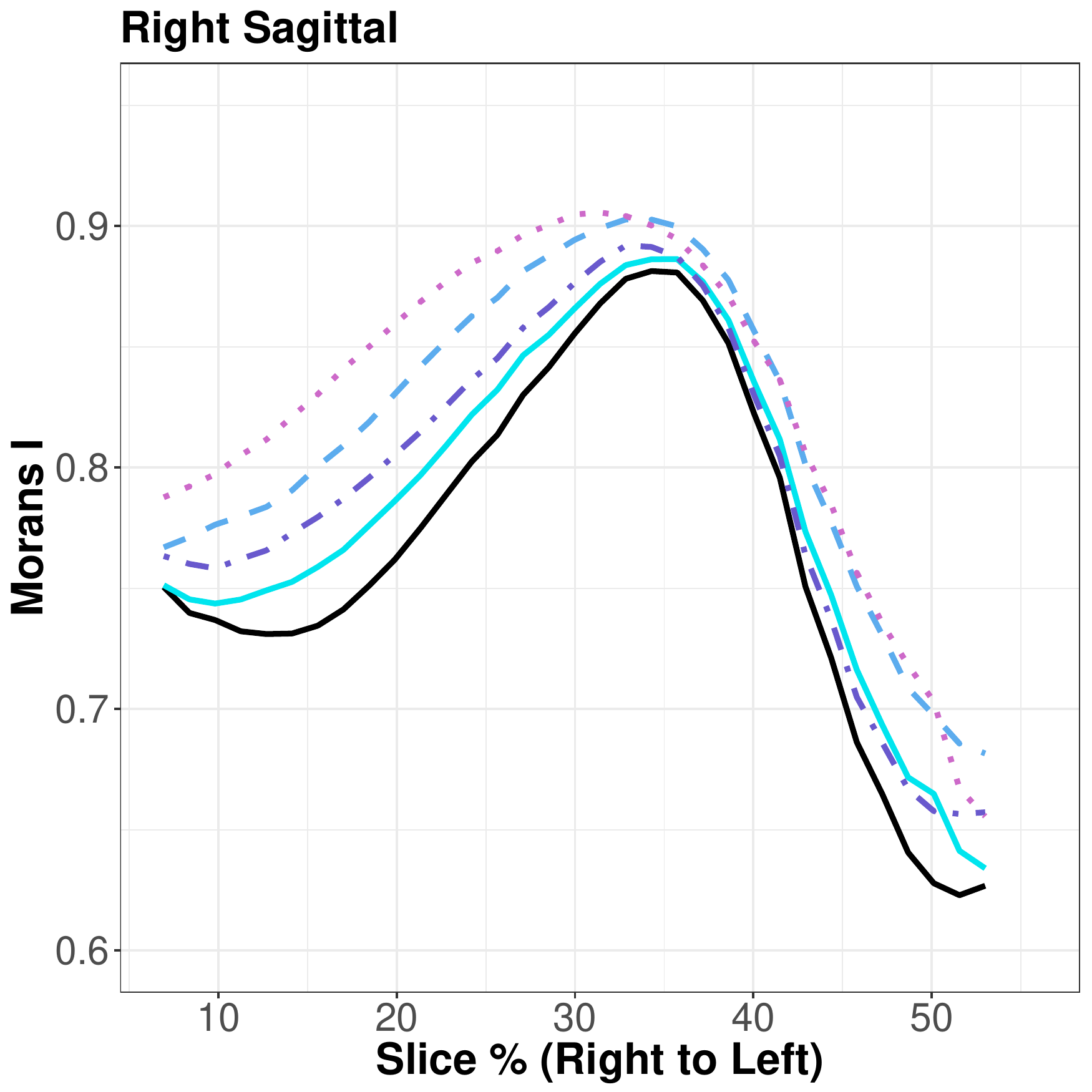}
  \includegraphics[width=2.25in]{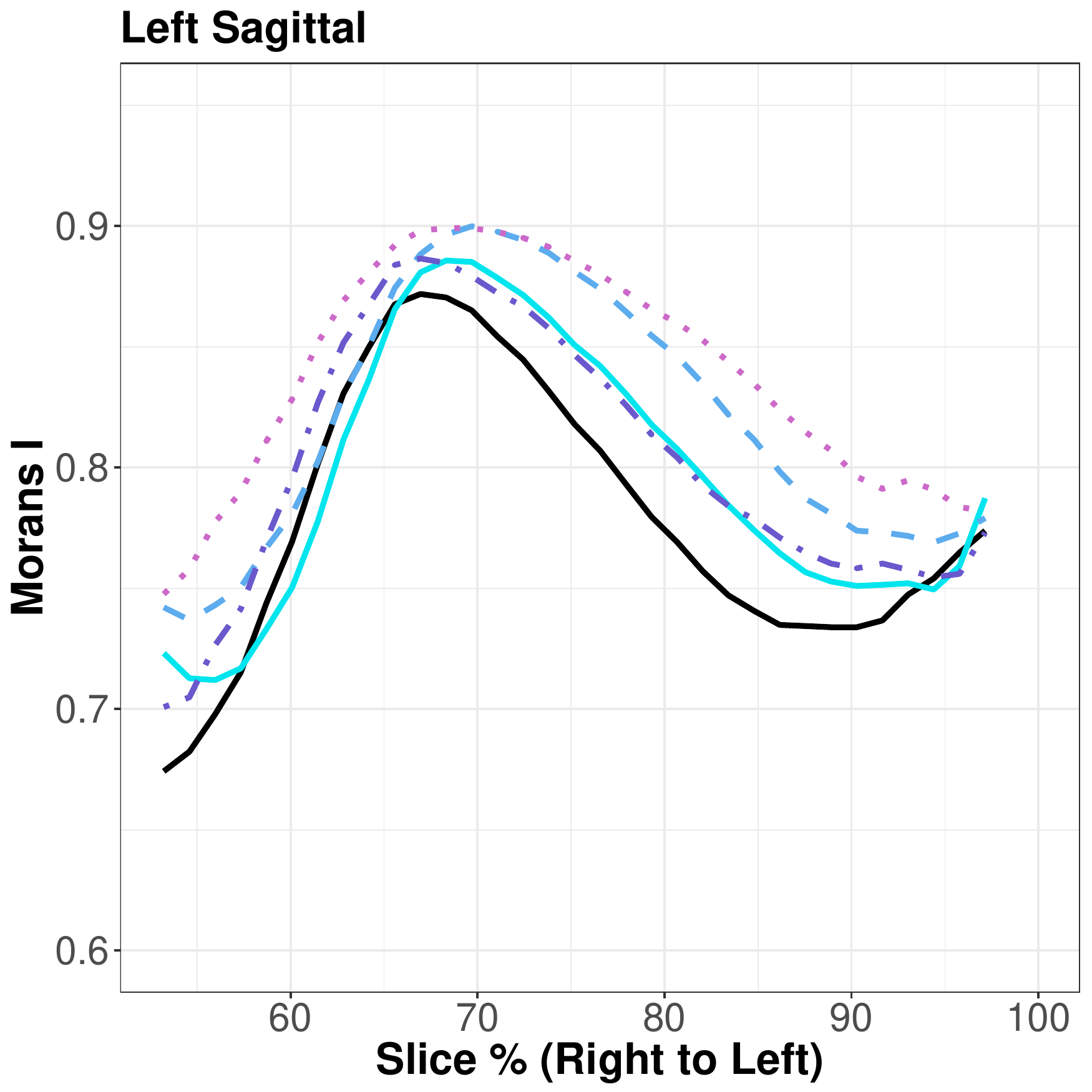}
  \caption[Crude:  Moran's $\mathcal{I}$ throughout the lungs by Scadding stage]{\linespread{1.3}\selectfont{}  Moran's $\mathcal{I}$ (8NN) throughout the lung by Scadding stage. The black solid lines represent the mean estimate for healthy controls; Stage I (solid light blue); Stage II (dashed blue); Stage III (dash-dotted purple); Stage IV (dotted pink).}
  \label{fig:MI1}
\end{figure}

 Moran's $\mathcal{I}$ shows a trend across healthy controls and stages of sarcoidosis (Figure~\ref{fig:MI1}). In general, healthy subjects have the lowest  Moran's $\mathcal{I}$, and sarcoidosis subjects classified as stage IV have the highest  Moran's $\mathcal{I}$, as we would expect. The trend in the stages from lowest to highest  Moran's $\mathcal{I}$ is as follows: Stage I and Stage III, Stage II, and Stage IV. This trend is most apparent in the top axial, middle coronal, and outer left/right sagittal regions. 

\subsection{Geary's $\mathcal{C}$}
\begin{figure}[H]
\centering
  \includegraphics[width=2.25in]{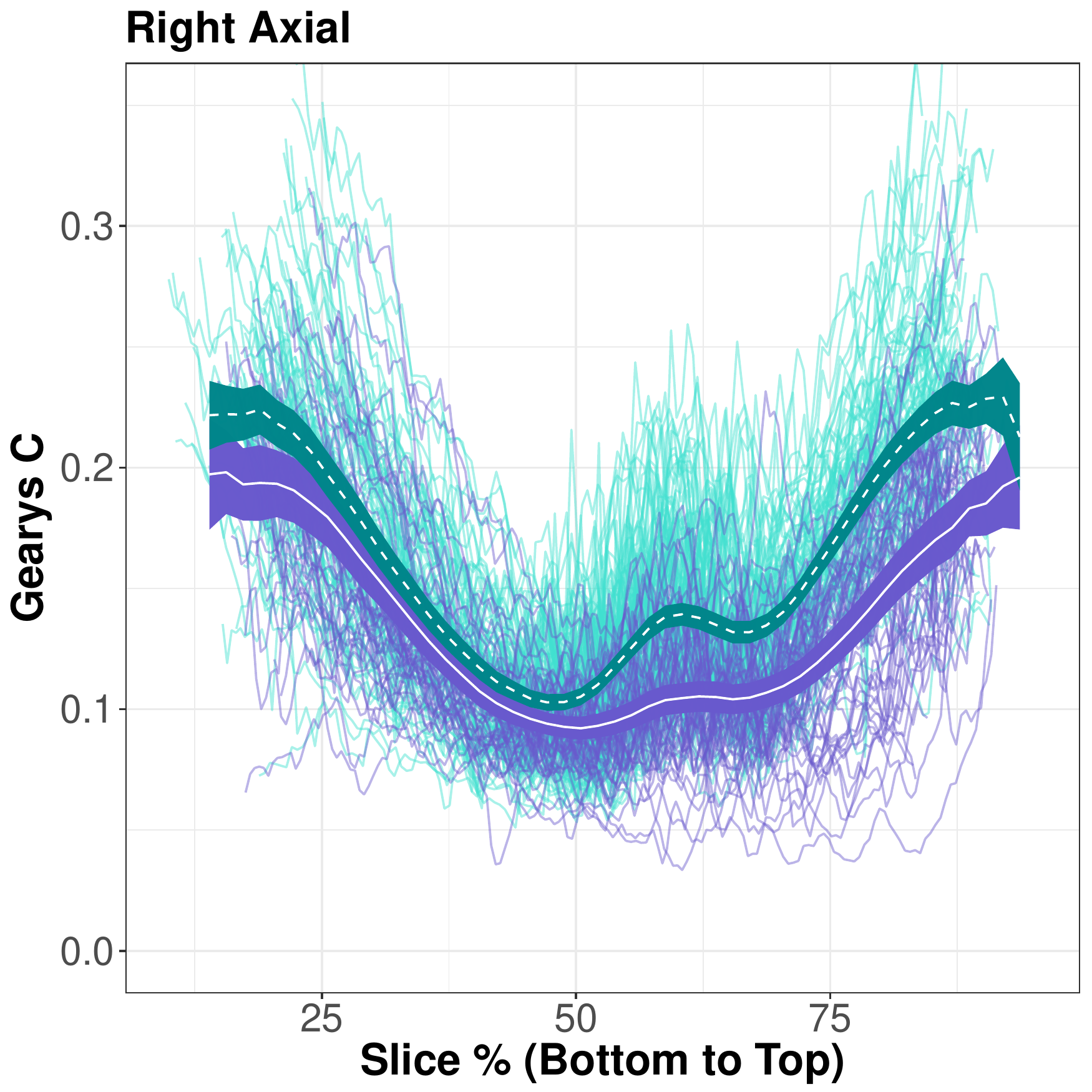}
  \includegraphics[width=2.25in]{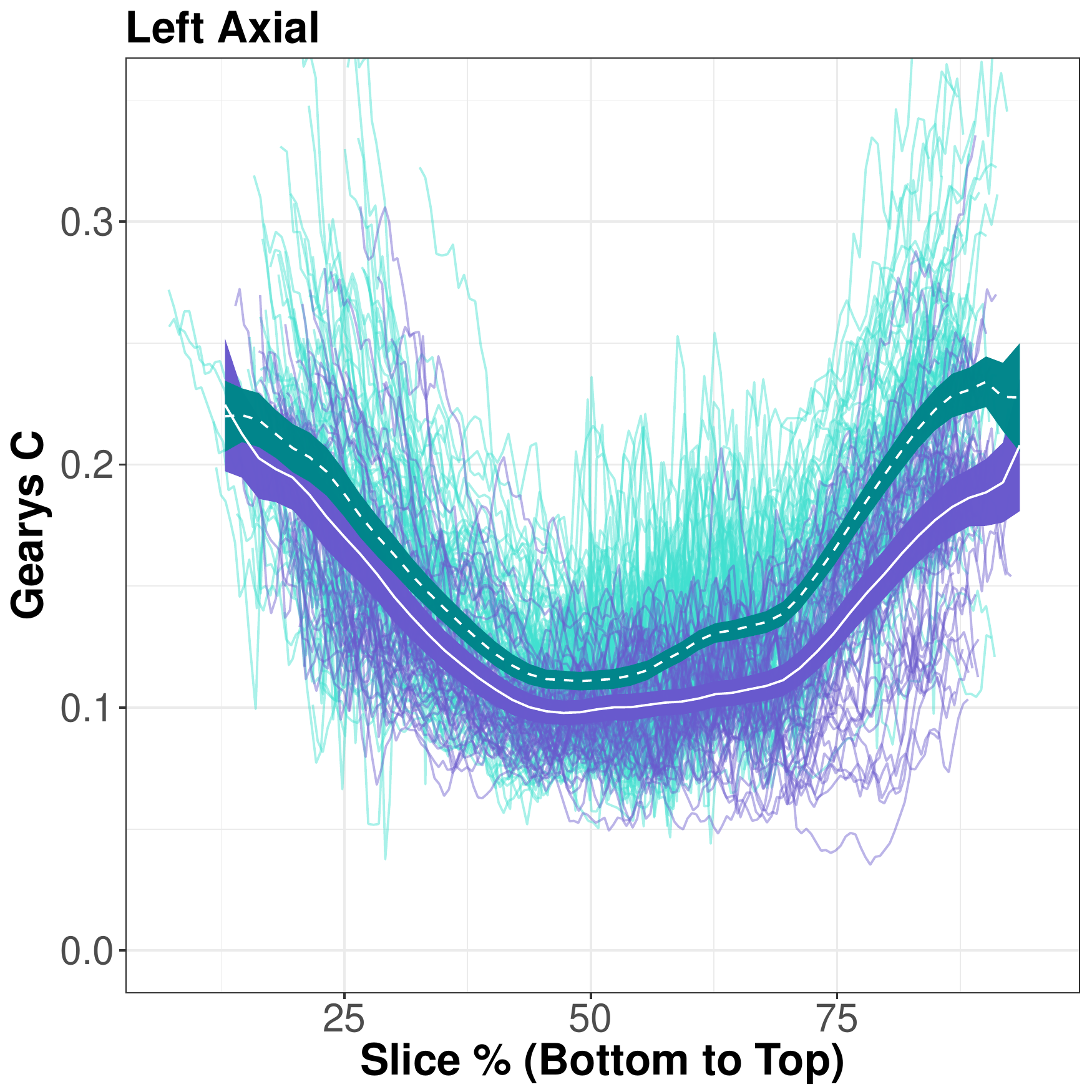}
  \includegraphics[width=2.25in]{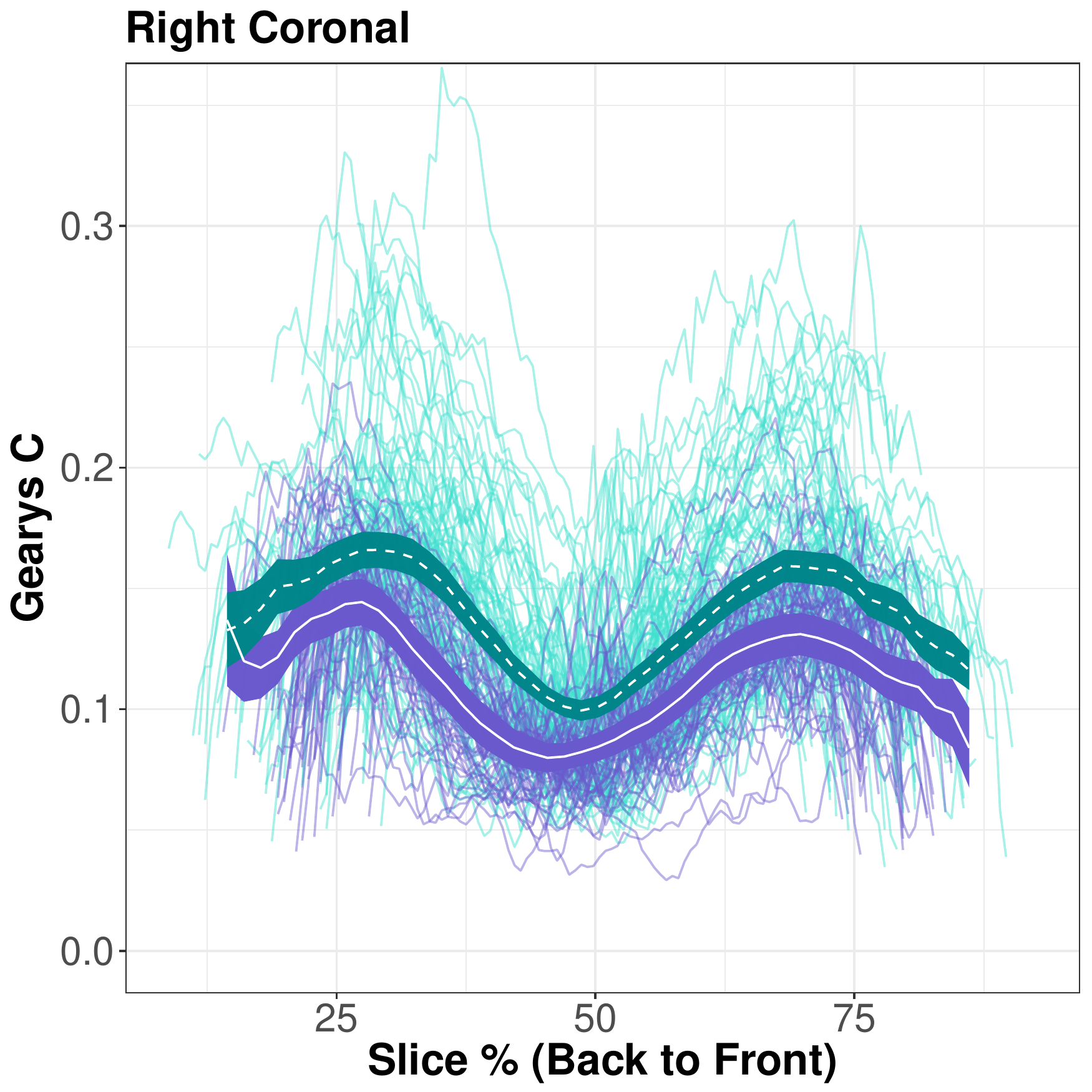}
  \includegraphics[width=2.25in]{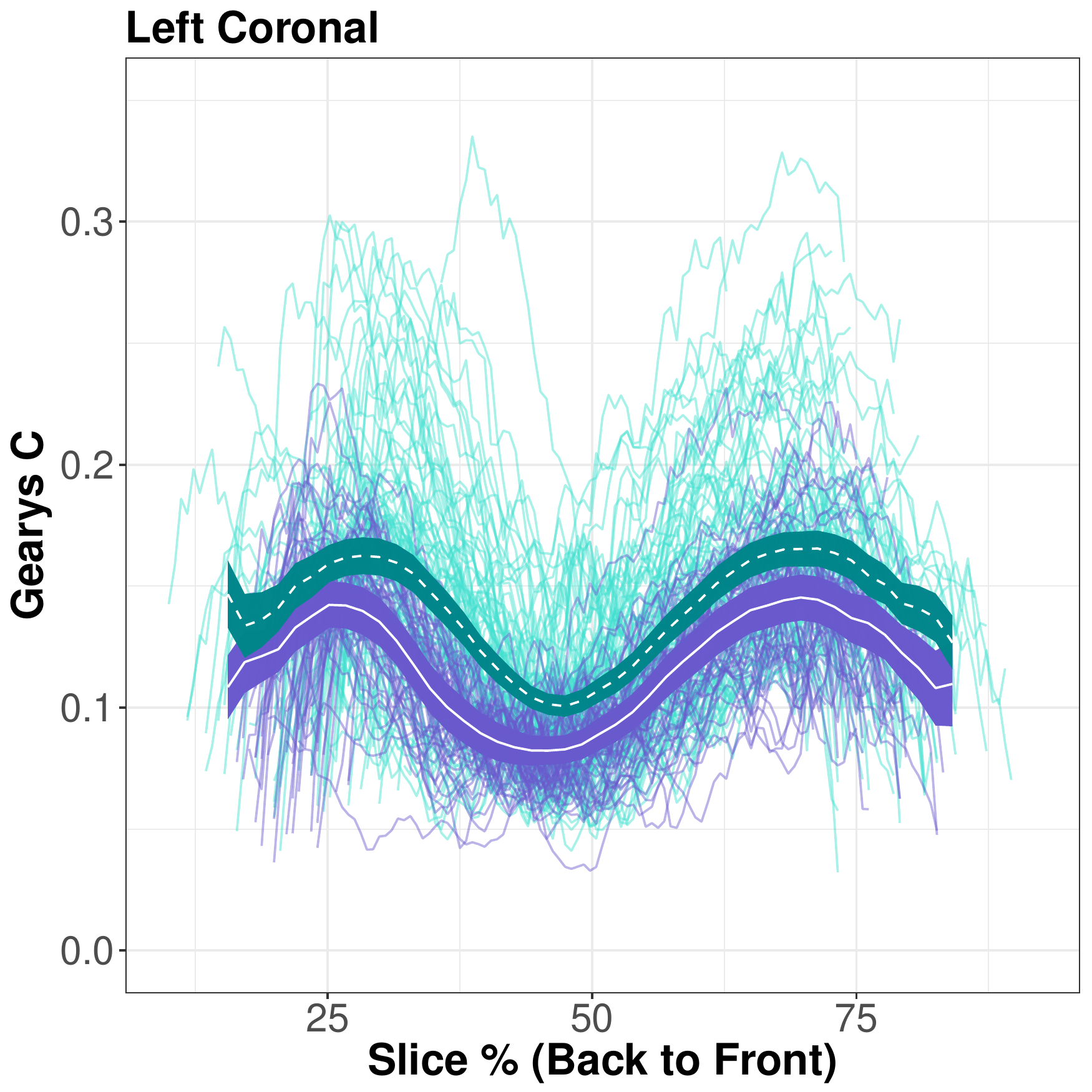}
  \includegraphics[width=2.25in]{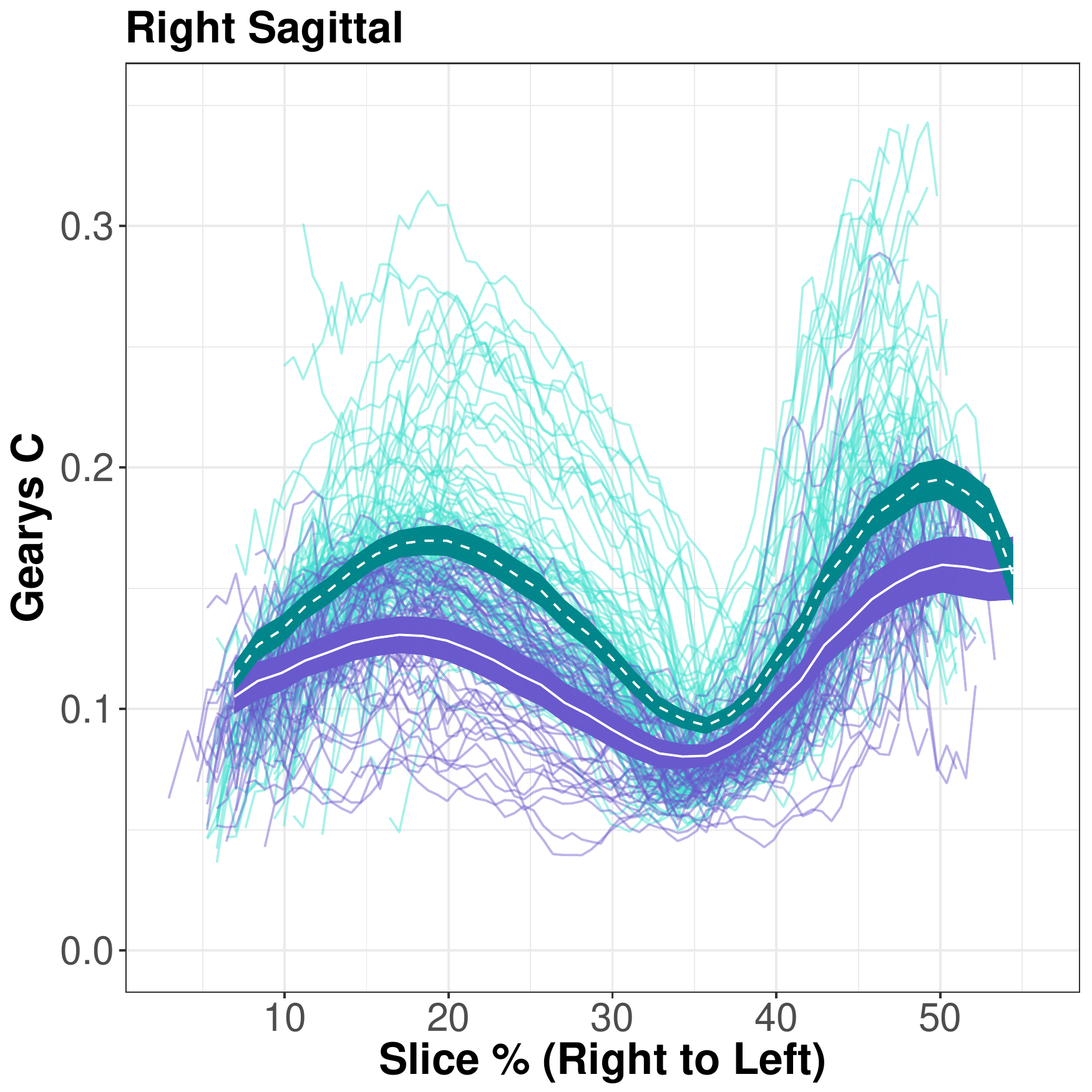}
  \includegraphics[width=2.25in]{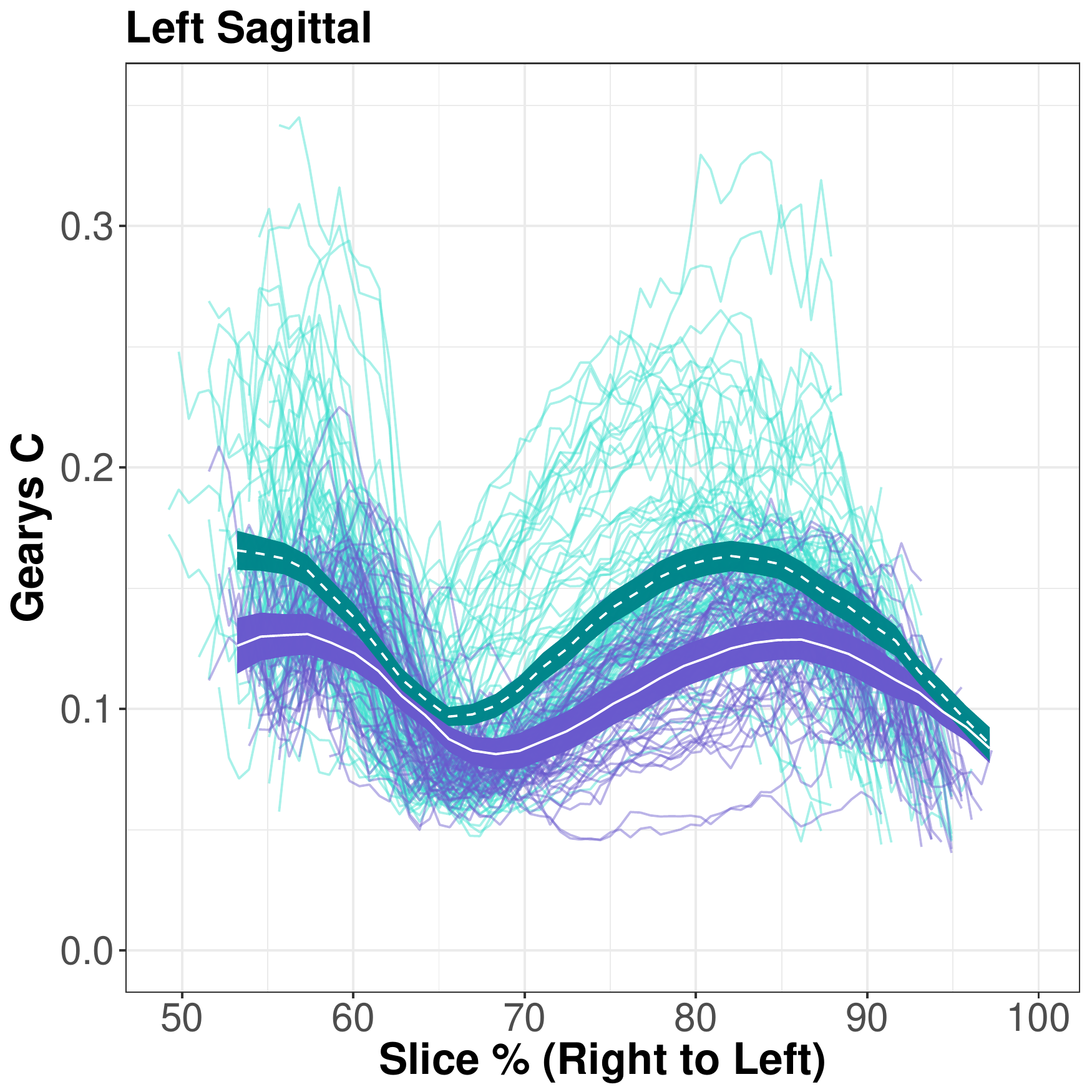}
  \caption[Crude: Geary's C throughout the lungs]{\linespread{1.3}\selectfont{} Geary's C throughout the lung. Bands represent 95\% CIs.  The green dashed lines indicate healthy controls, and the purple solid lines represent subjects with sarcoidosis.}
  \label{fig:gc}
\end{figure}

For all subjects, Geary's C ranges from 0.03 to 0.42 units, indicative of positive spatial structure. In the axial plane, Geary's C is lowest in the middle of the lung, and highest at the bottom and top of the lung. In the coronal plane, Geary's C is lowest in the middle of the lung, and highest at the back and front of the lung. In the sagittal plane, Geary's C is lowest on the middle of the right and left of the lungs, and highest at the inner part (towards the center of the body) (Figure~\ref{fig:gc}). These results indicate that the lung shows the most local spatial structure in the middle axial, coronal, and sagittal regions, and the least spatial structure in the bottom/top axial, back/front coronal, and inner sagittal regions. 

Without adjusting for covariates (Figure~\ref{fig:gc}), Geary's C significantly differentiates between subjects with and without sarcoidosis. The areas with the most differentiation between diseased and control subjects are the top axial, middle coronal, and outer sagittal regions. In general, subjects with sarcoidosis have lower Geary's C than healthy controls, indicative of more local, positive spatial structures. 

\begin{figure}[H]
\centering
  \includegraphics[width=2.25in]{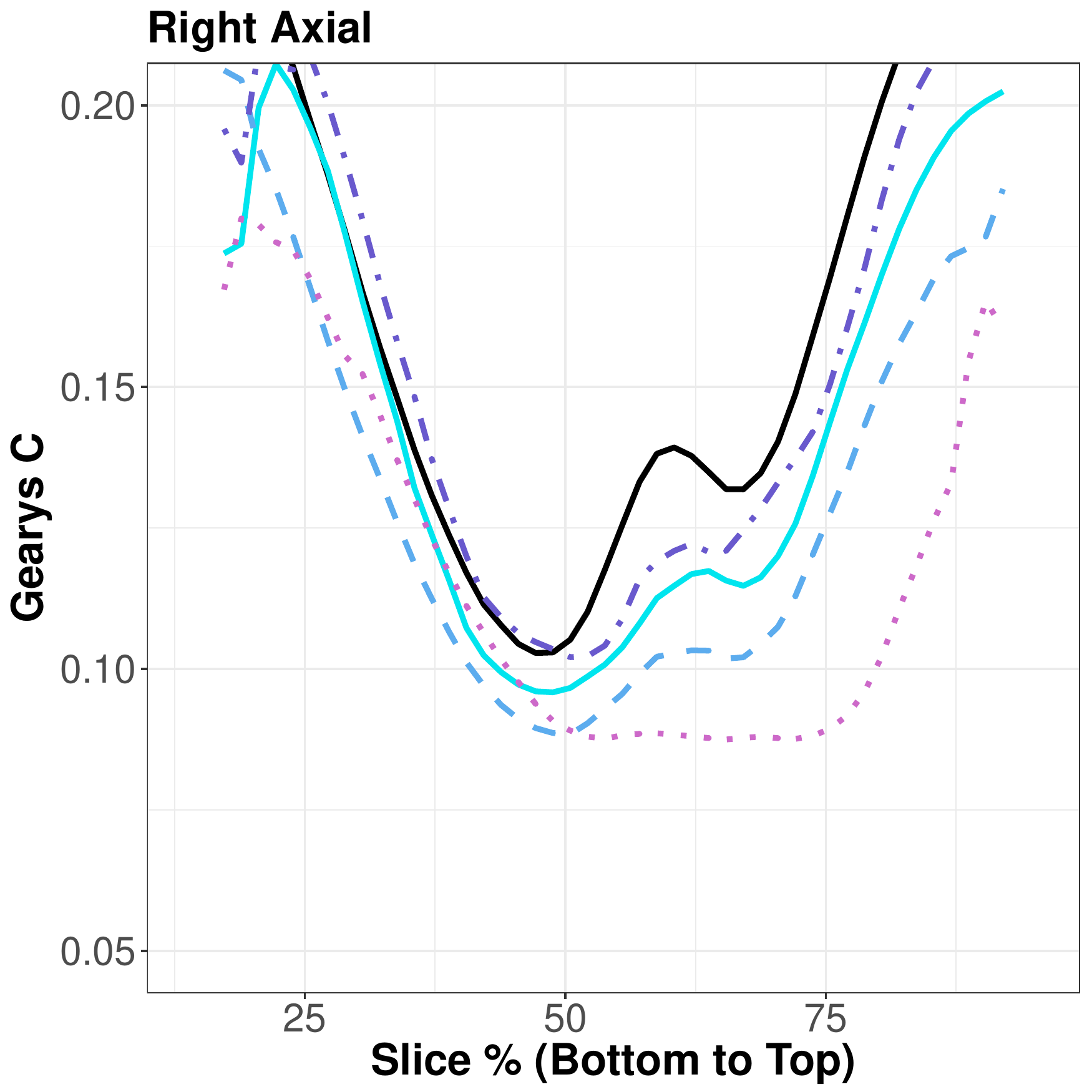}
  \includegraphics[width=2.25in]{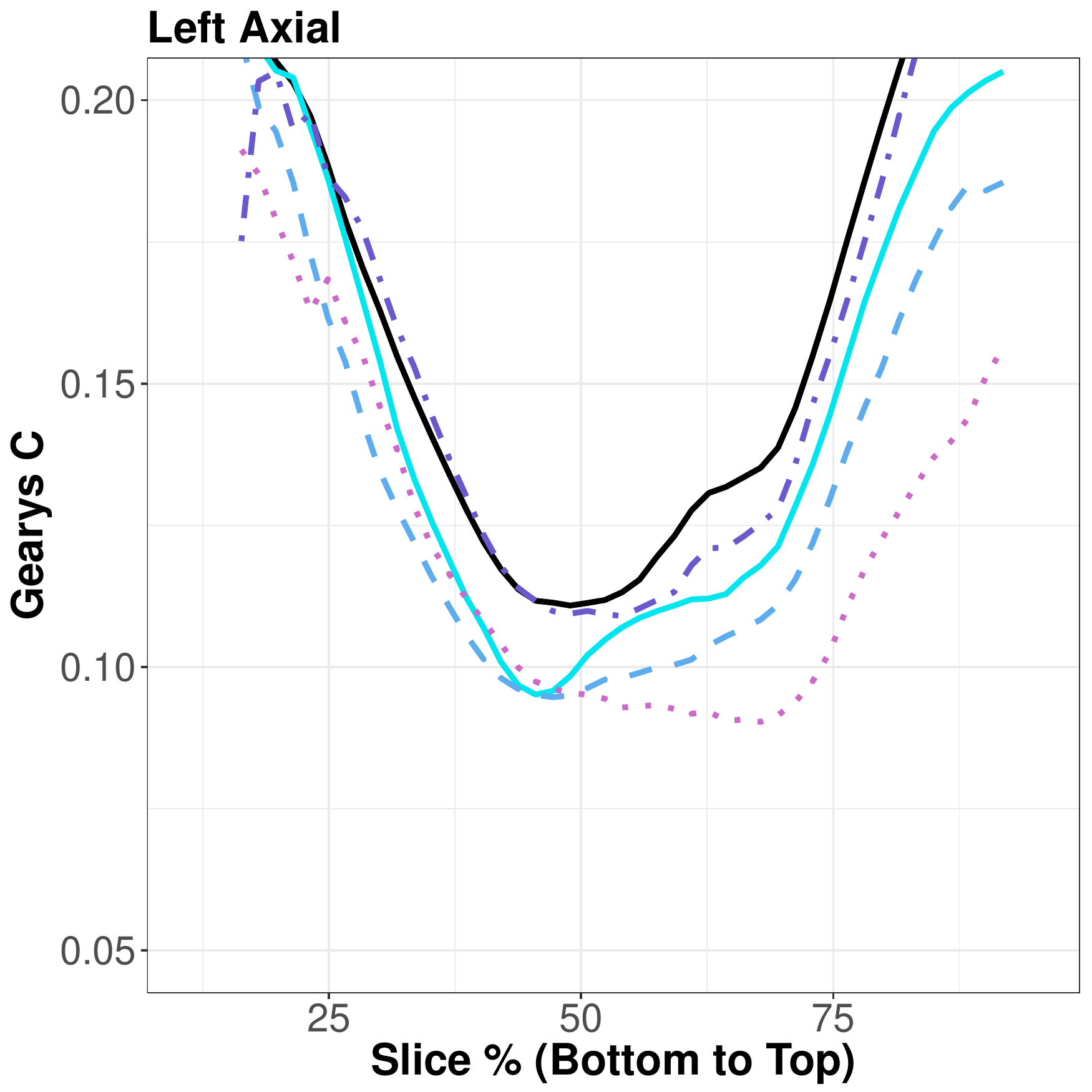}
  \includegraphics[width=2.25in]{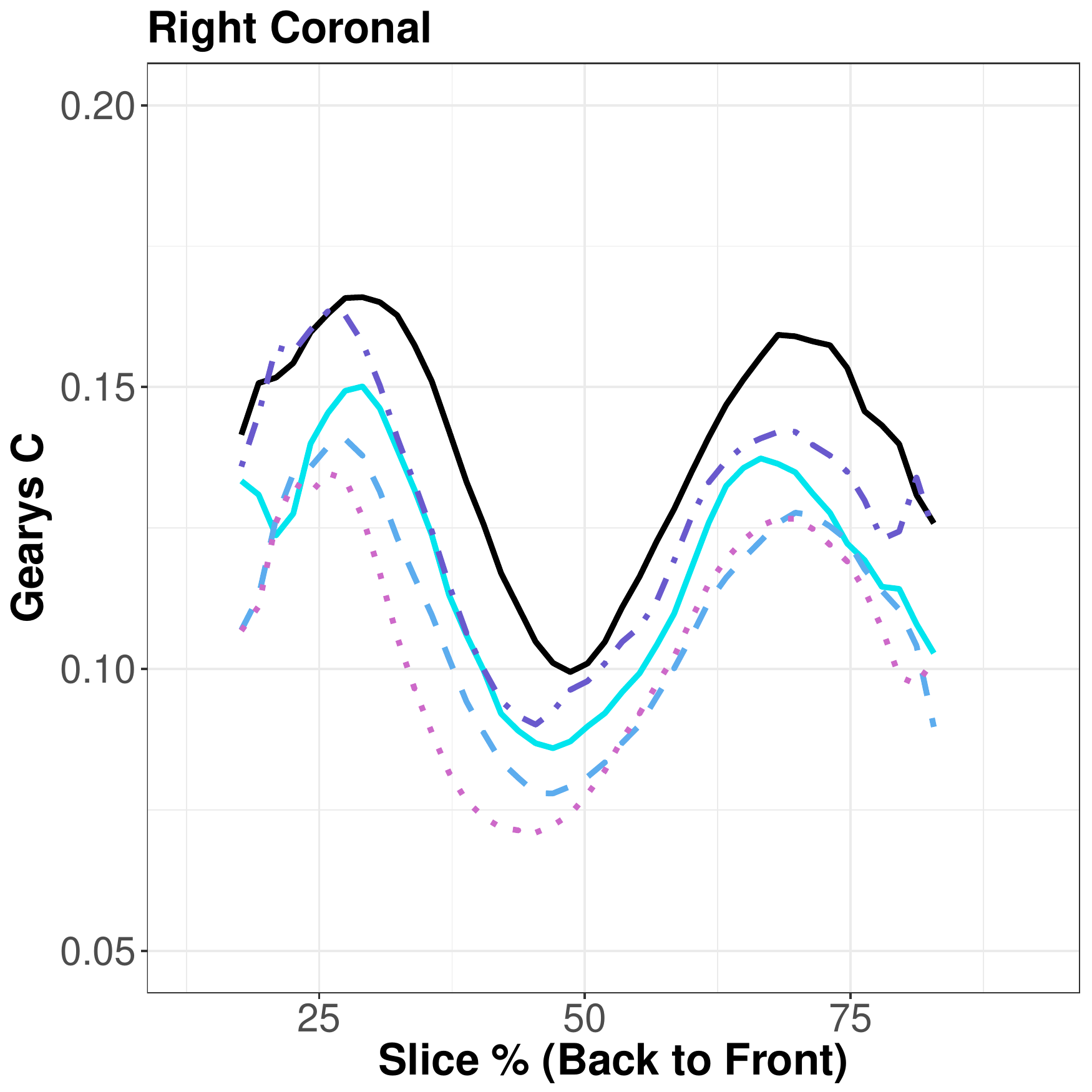}
  \includegraphics[width=2.25in]{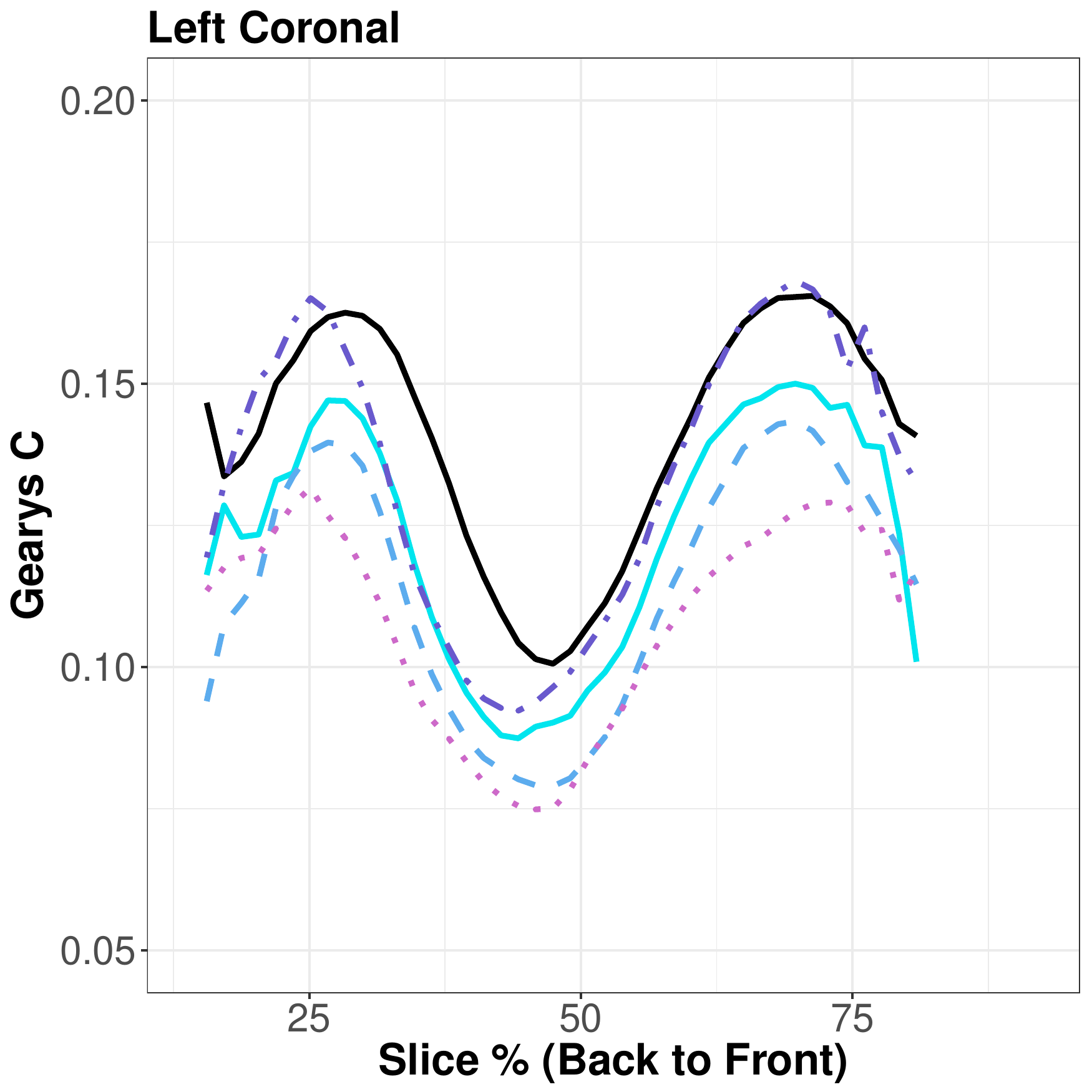}
  \includegraphics[width=2.25in]{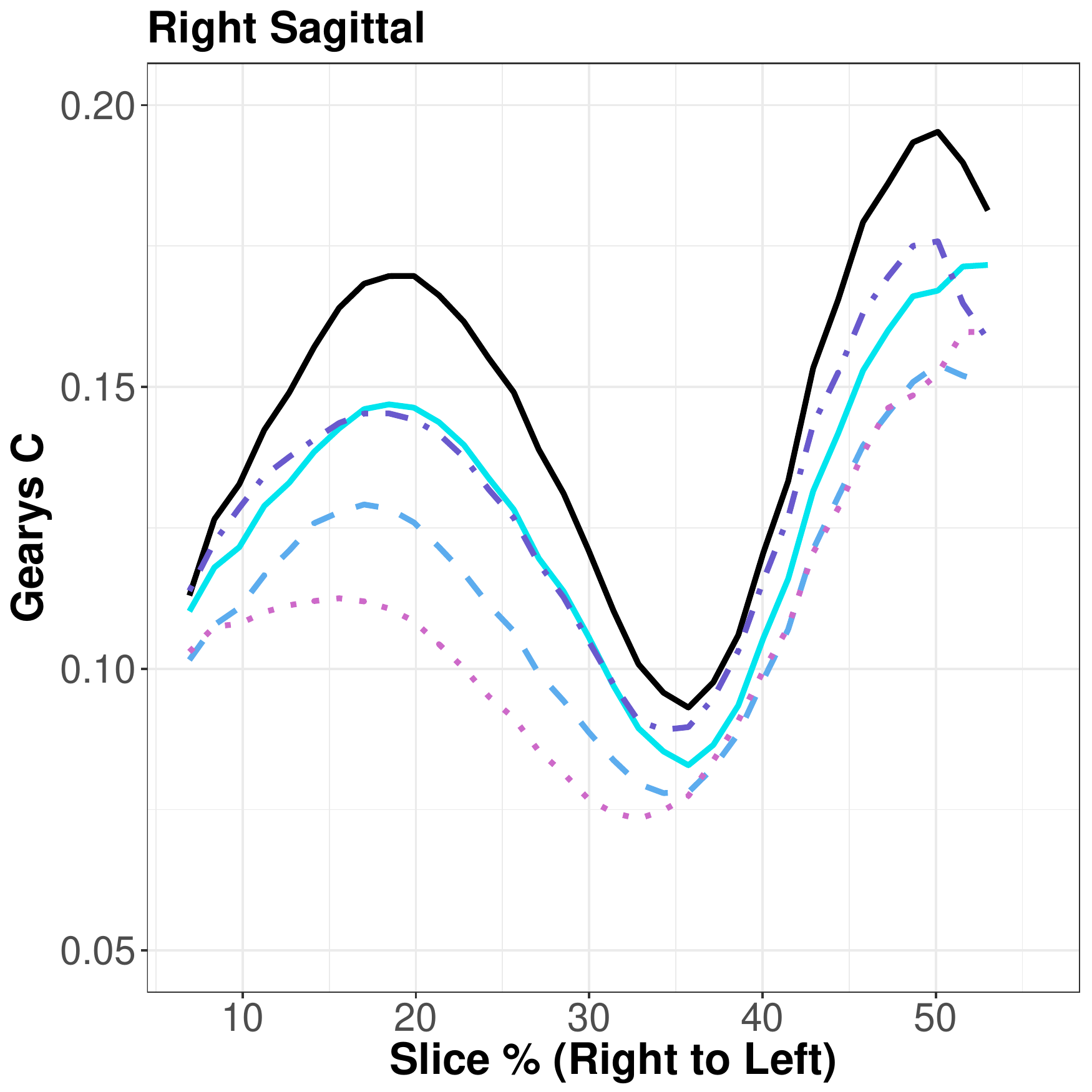}
  \includegraphics[width=2.25in]{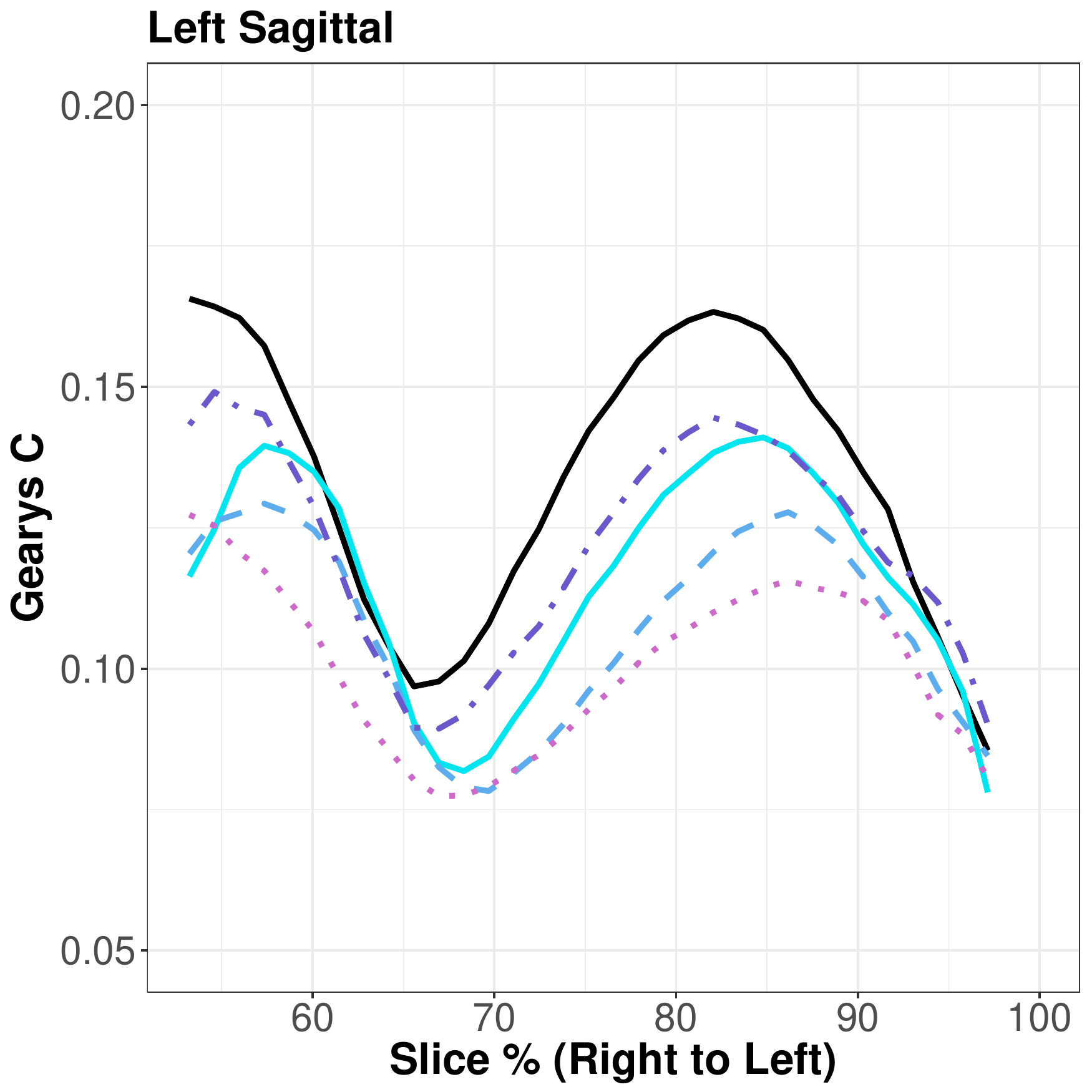}
  \caption[Crude: Geary's C throughout the lungs by Scadding stage]{\linespread{1.3}\selectfont{} Geary's C throughout the lung by Scadding stage. The black solid lines represent the mean estimate for healthy controls; Stage I (solid light blue); Stage II (dashed blue); Stage III (dash-dotted purple); Stage IV (dotted pink).}
  \label{fig:GC1}
\end{figure}

Geary's C also shows a trend across healthy controls and stages of sarcoidosis (Figure~\ref{fig:GC1}). In general, healthy subjects have the highest Geary's C, and sarcoidosis subjects classified as stage IV have the lowest, as we would expect. Similar to  Moran's $\mathcal{I}$, the trend in the stages from highest to lowest Geary's C is as follows: Stage I and Stage III, Stage II, and Stage IV. This trend is most apparent in the top axial, middle coronal, and outer left/right sagittal regions.

\subsection{Fractal Dimension, $\mathcal{D}$}
\begin{figure}[H]
\centering
  \includegraphics[width=2.25in]{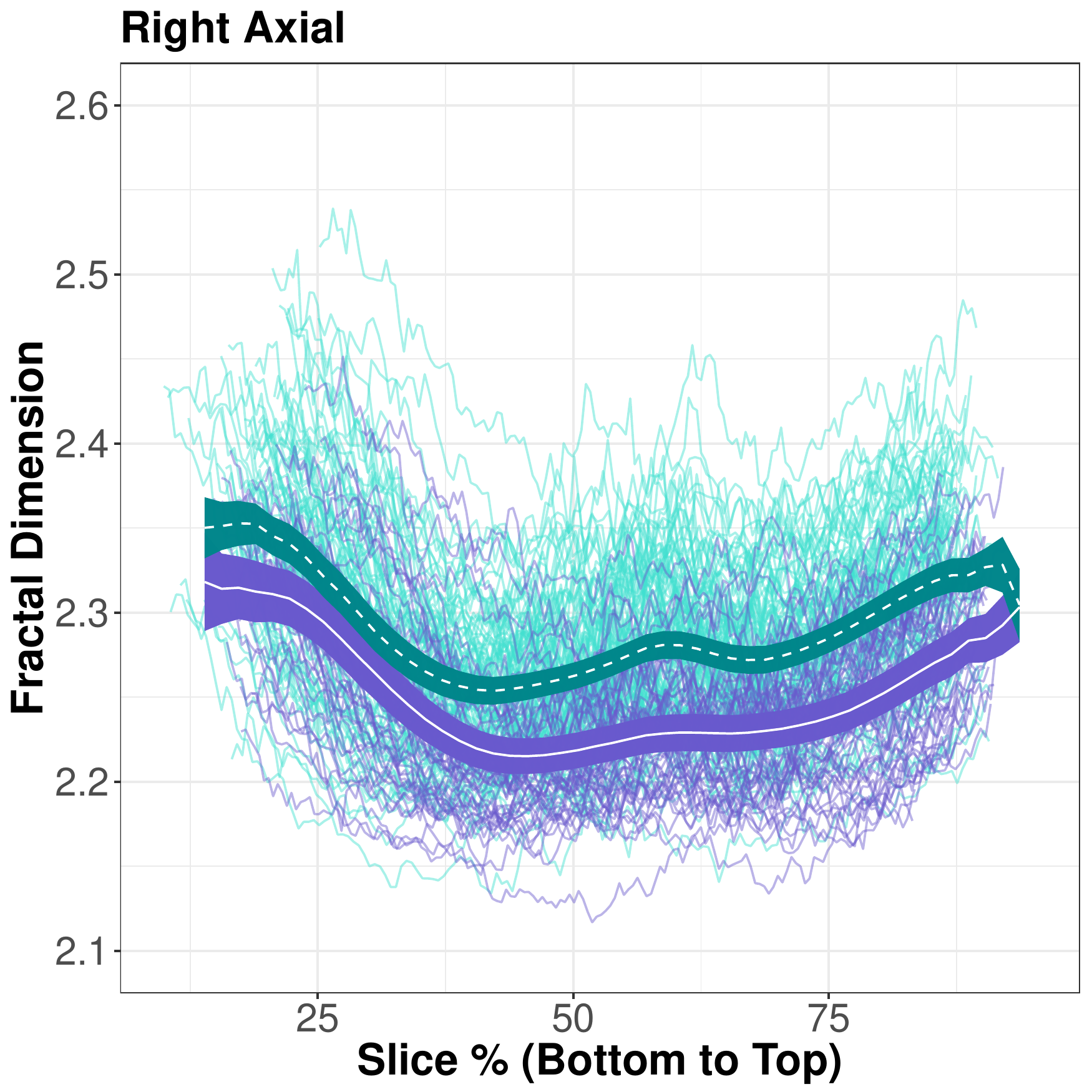}
  \includegraphics[width=2.25in]{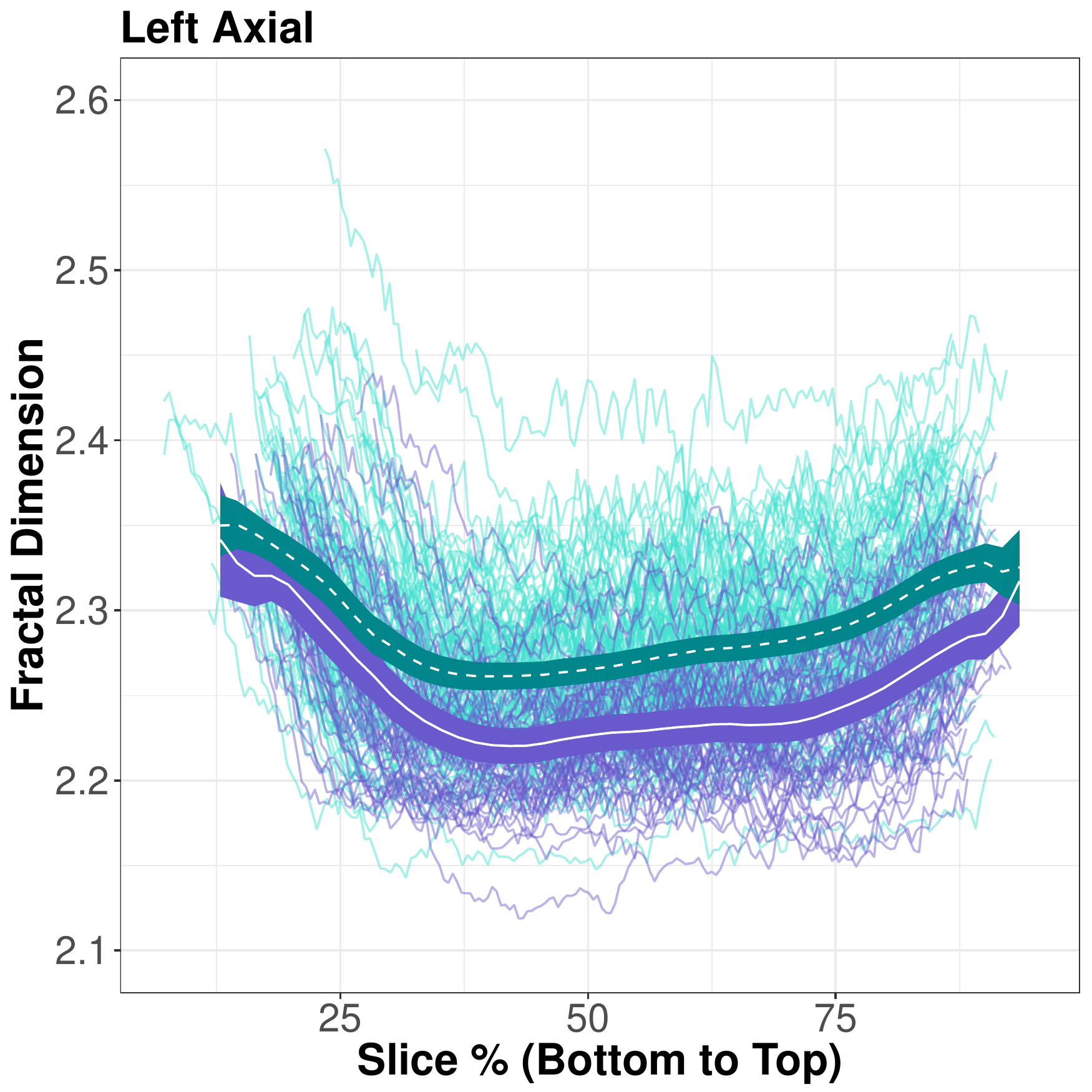}
  \includegraphics[width=2.25in]{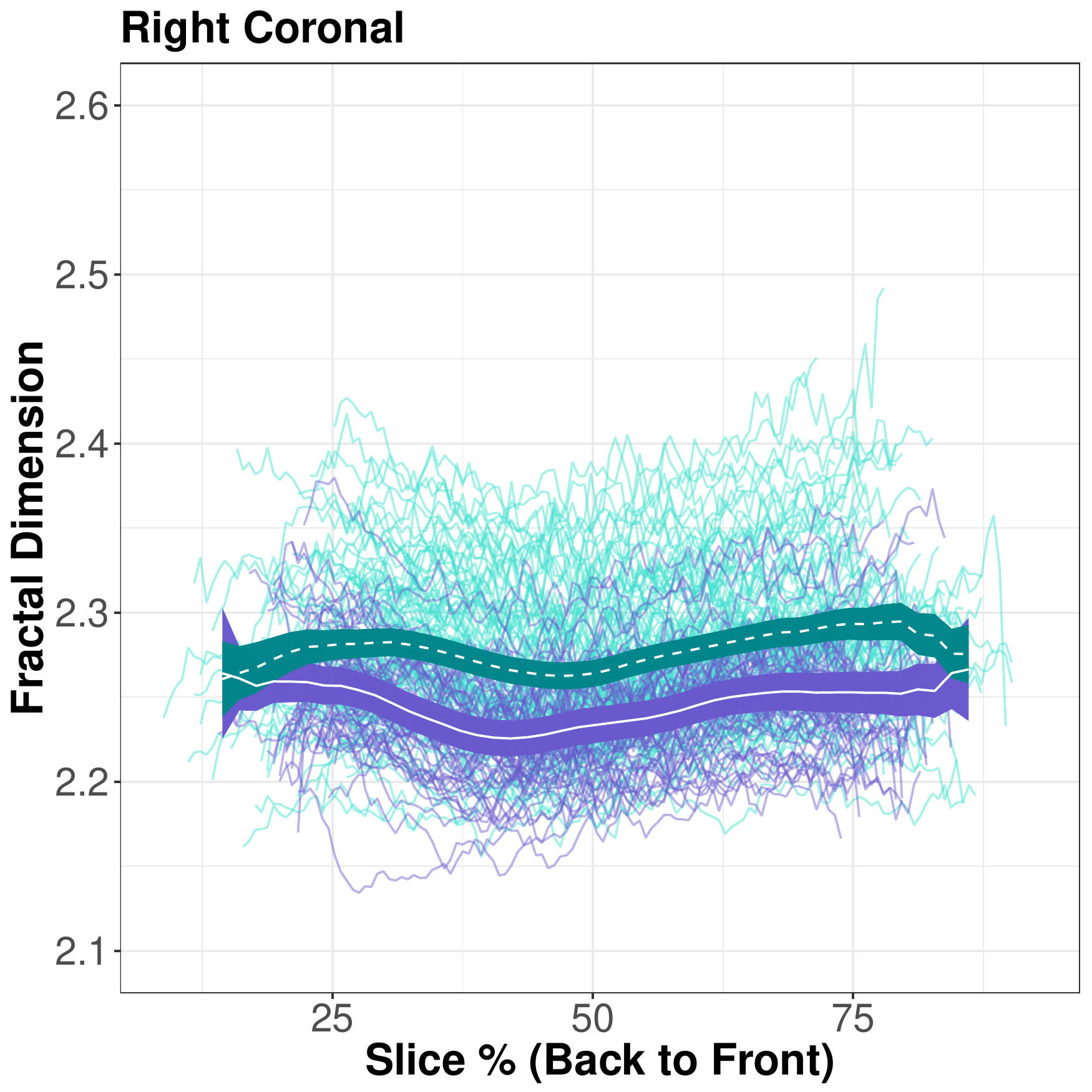}
  \includegraphics[width=2.25in]{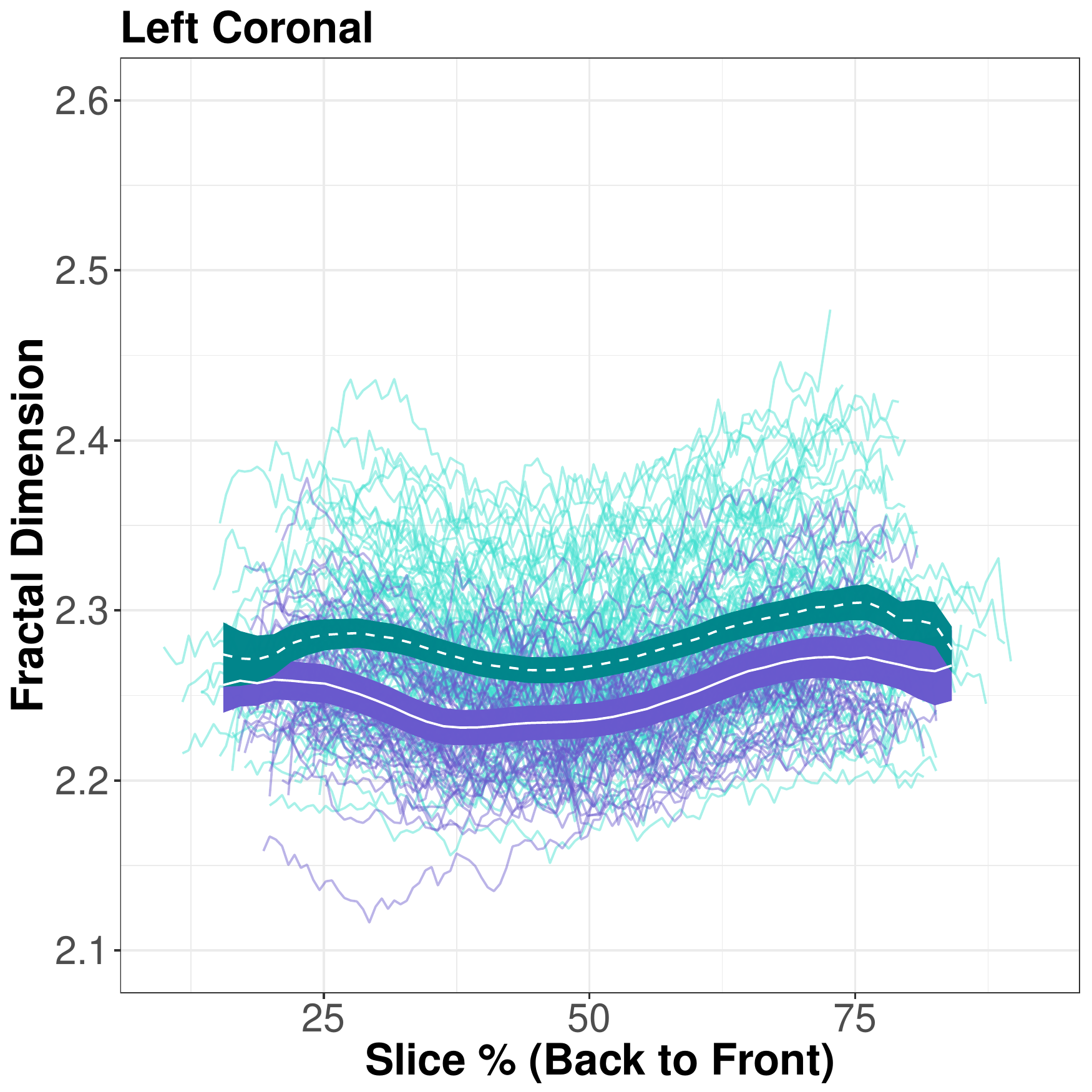}
  \includegraphics[width=2.25in]{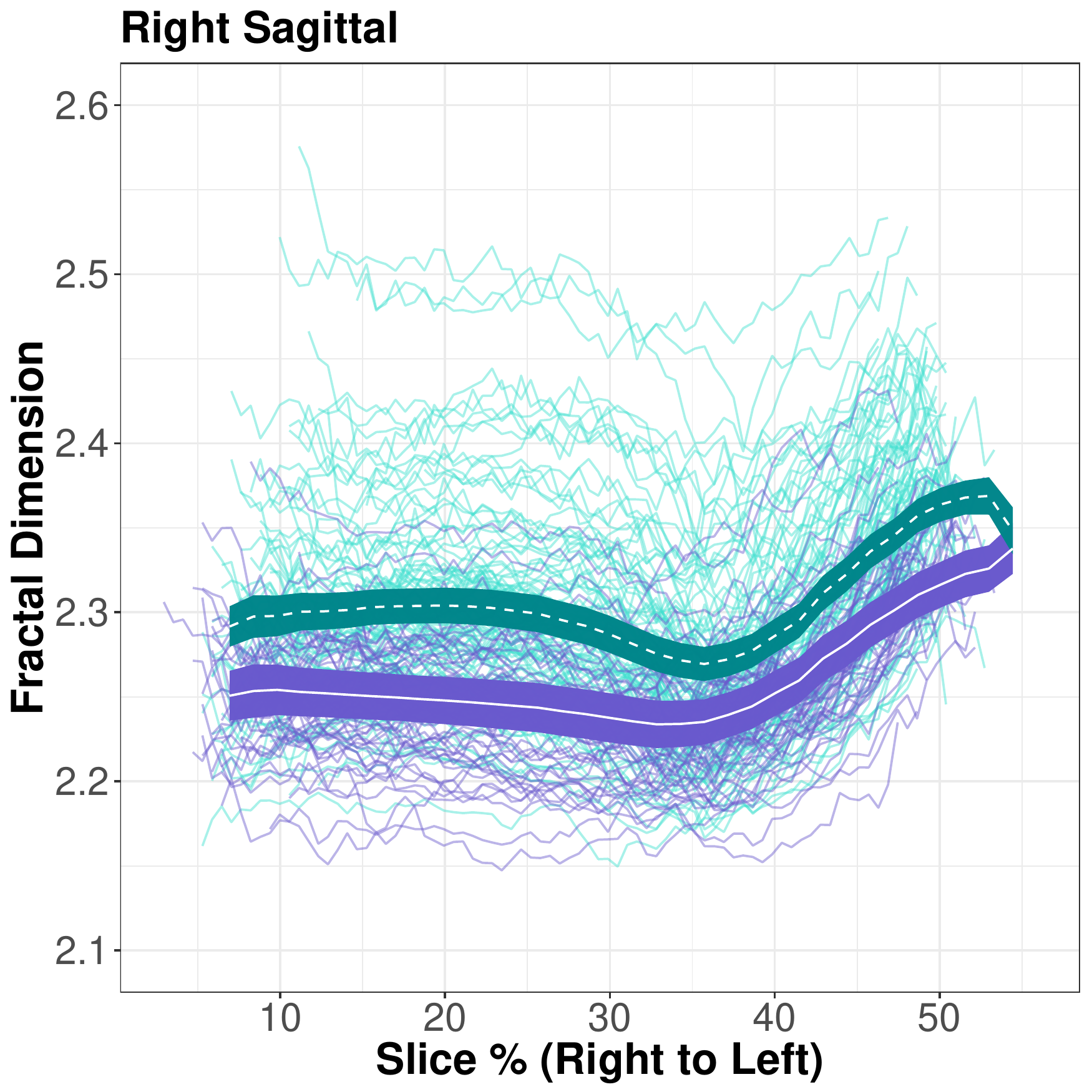}
  \includegraphics[width=2.25in]{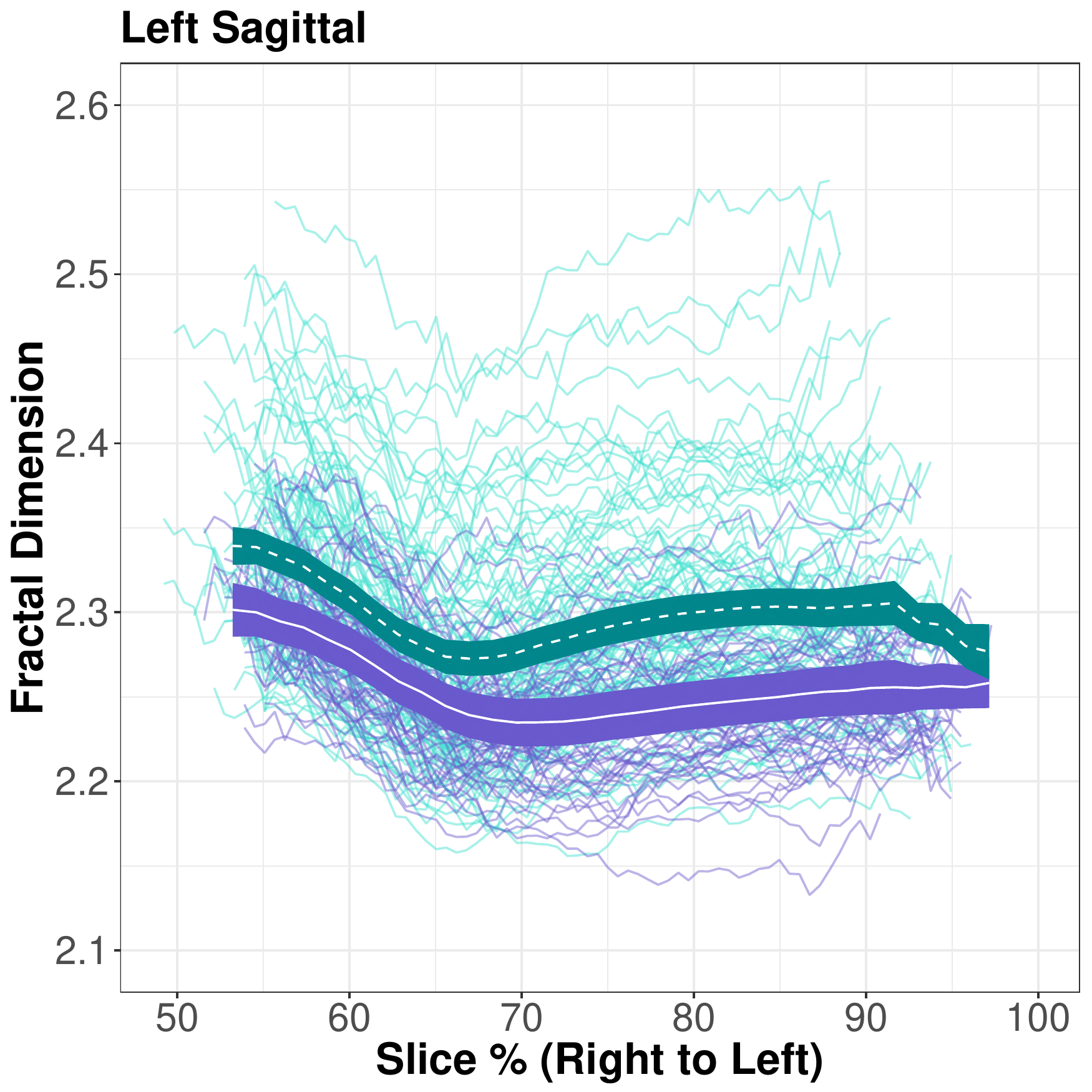}
  \caption[Crude: Fractal dimension throughout the lungs]{\linespread{1.3}\selectfont{} Fractal dimension throughout the lung. Bands represent 95\% CIs. The green dashed lines indicate healthy controls, and the purple solid lines represent subjects with sarcoidosis.}
  \label{fig:fd}
\end{figure}

For all subjects, fractal dimension ranges from 2.12 to 2.61 units. In the axial plane, fractal dimension is highest at the bottom of the lung, and lowest in the middle of the lung. In the coronal plane, fractal dimension is fairly constant from the back slices to the front. In the sagittal plane, fractal dimension is highest in the middle of the lungs, and lowest on the outer right and left slices (Figure~\ref{fig:fd}). These results indicate that the lung is ``roughest" and most``self-similar" at the bottom axial and inner sagittal, and ``smoothest" and least ``self-similar" at the middle axial and outer sagittal regions. 

Fractal dimension significantly differentiates between subjects with and without sarcoidosis (Figure~\ref{fig:fd}). Subjects with sarcoidosis have significantly lower fractal dimension than healthy controls in nearly all parts of the lung, regardless of right/left lung and anatomical orientation. This is indicative of a lower degree of self-similarity and more smoothness in sarcoidosis subjects as compared to controls throughout the entire lung. 

\begin{figure}[H]
\centering
  \includegraphics[width=2.25in]{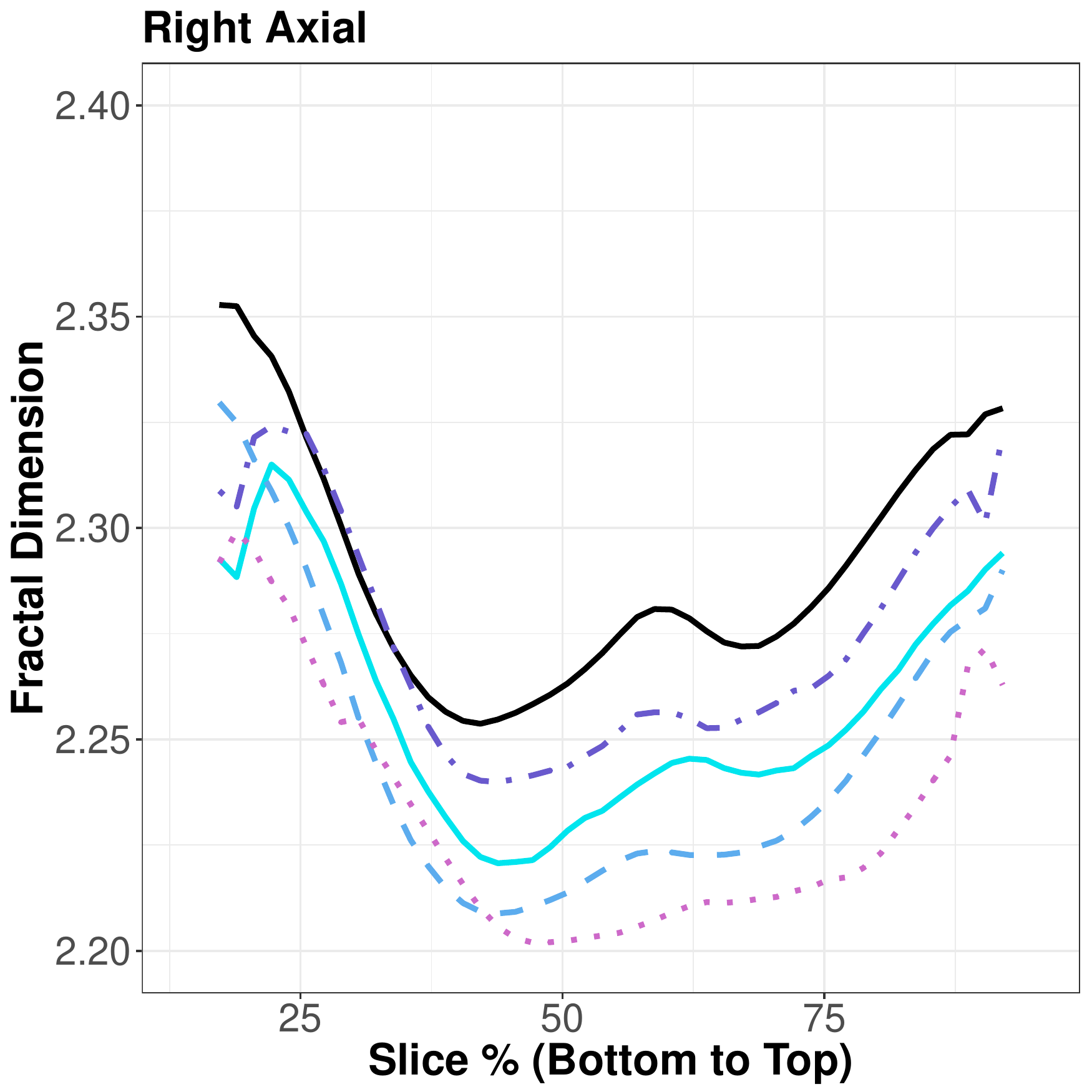}
  \includegraphics[width=2.25in]{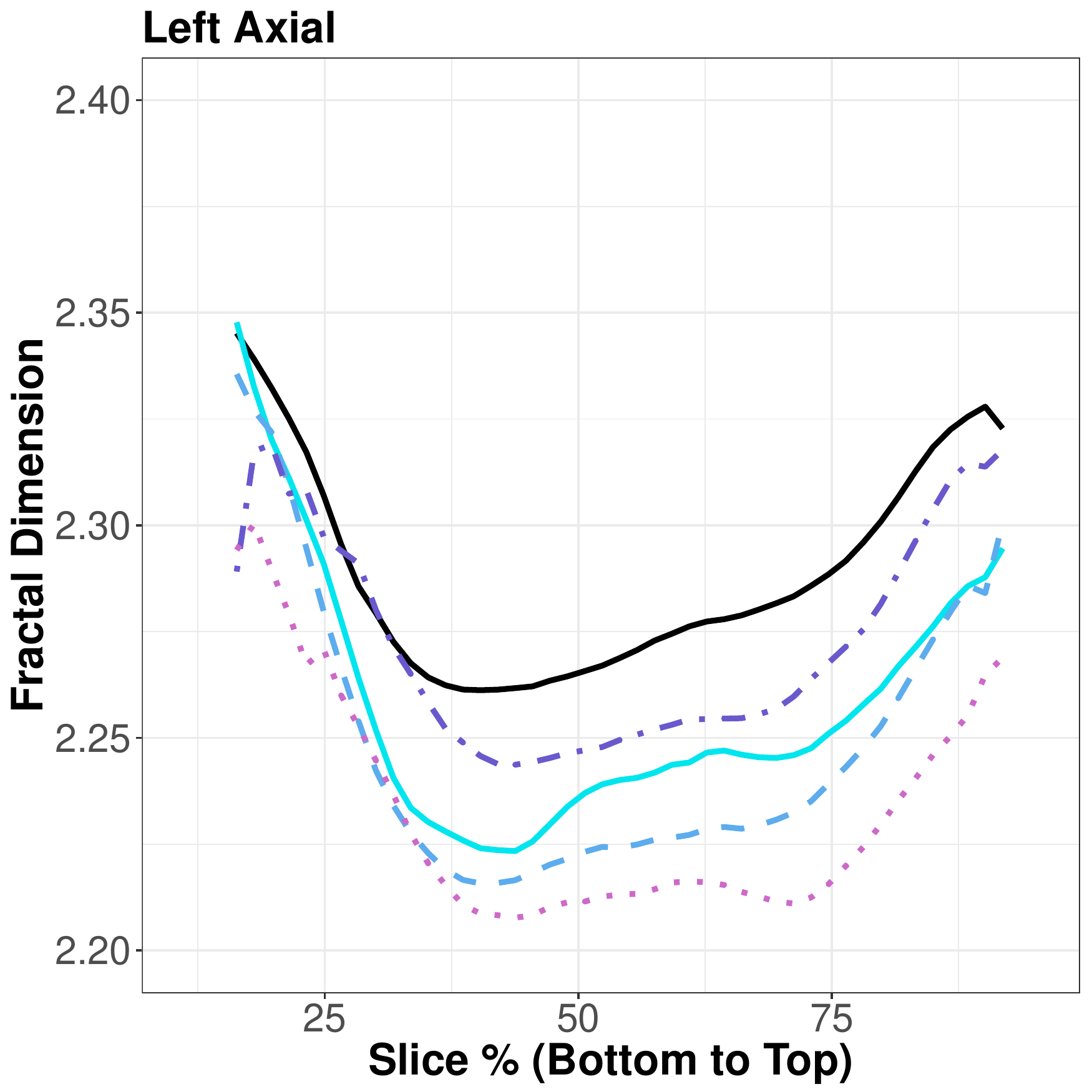}
  \includegraphics[width=2.25in]{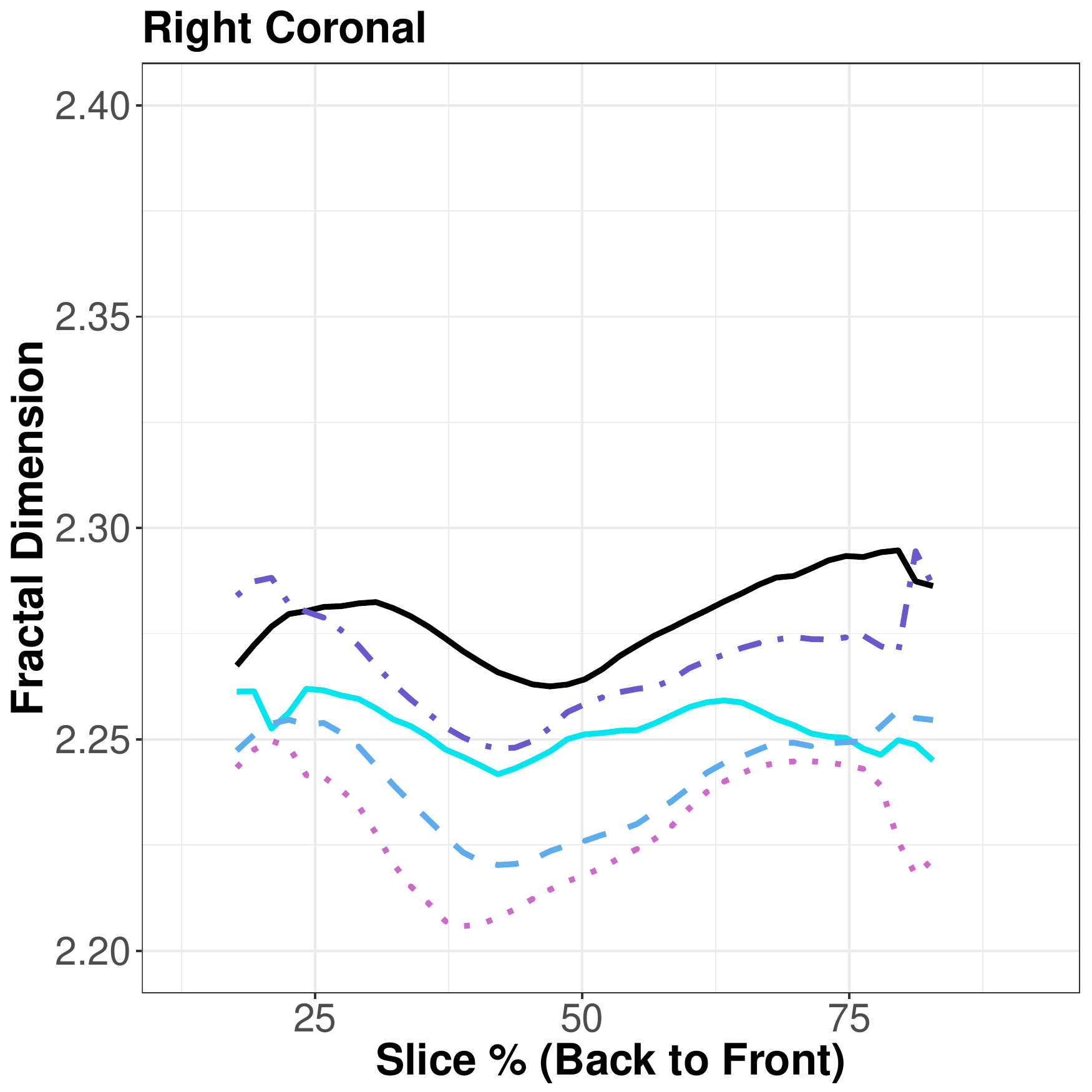}
  \includegraphics[width=2.25in]{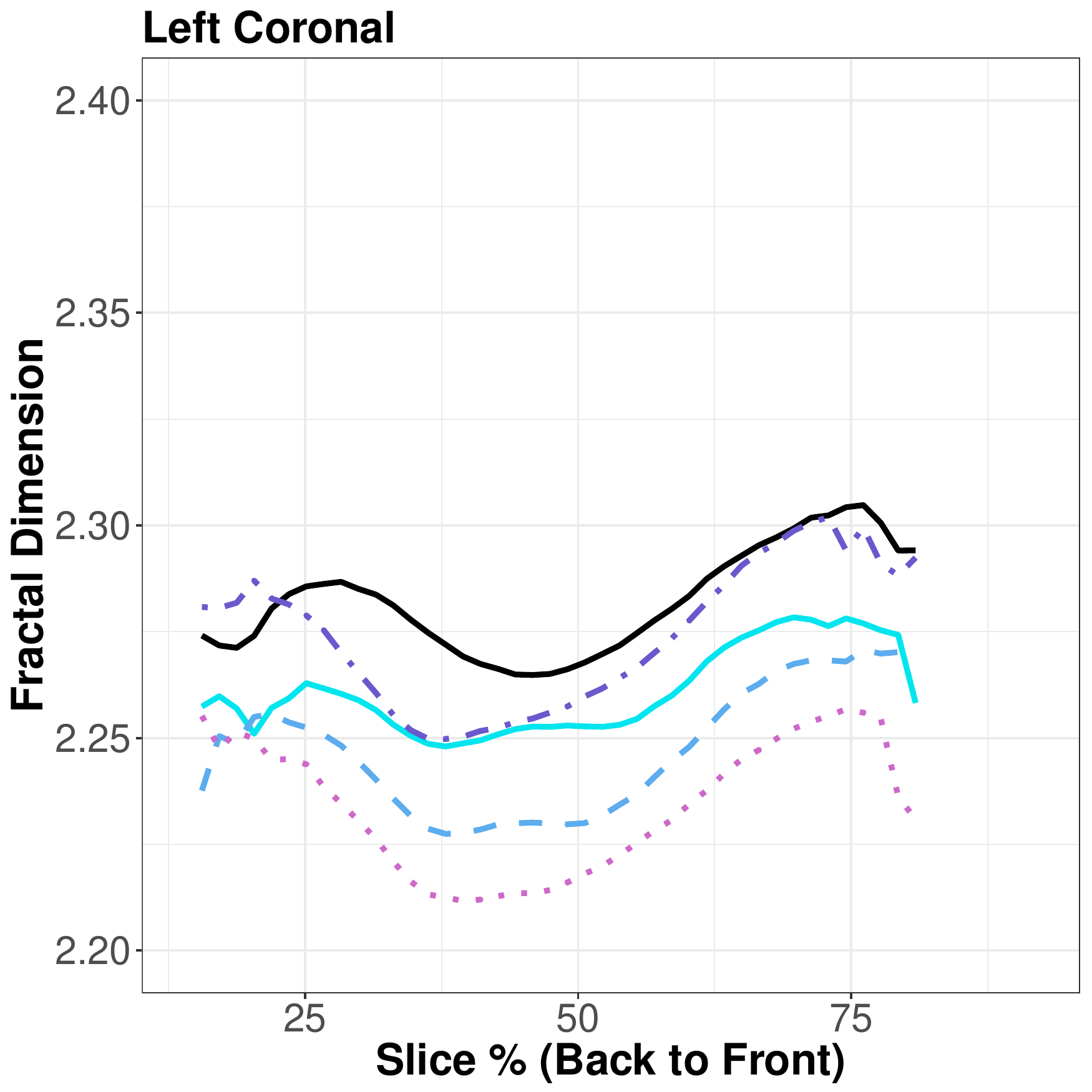}
  \includegraphics[width=2.25in]{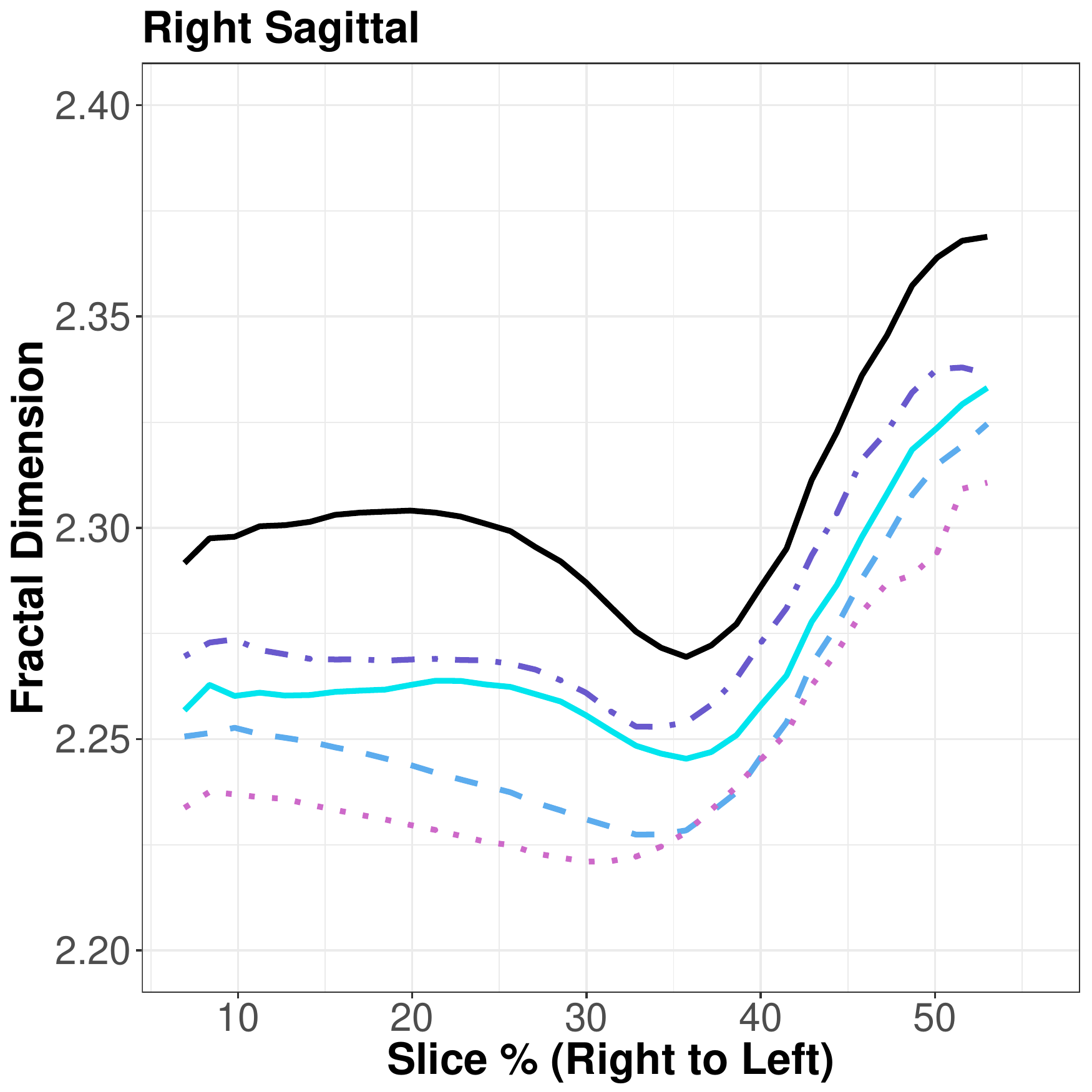}
  \includegraphics[width=2.25in]{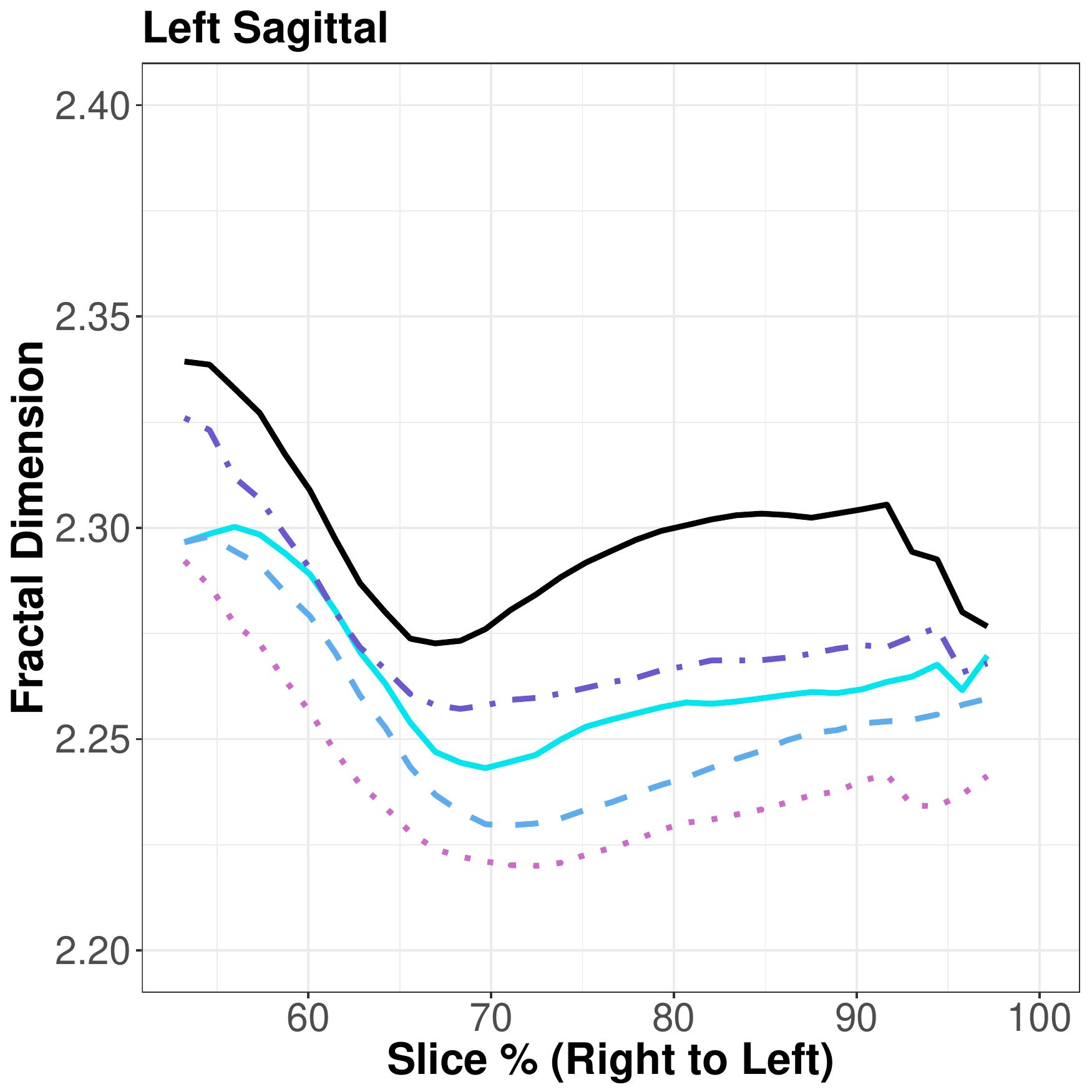}
  \caption[Crude: Fractal dimension throughout the lungs by Scadding stage]{\linespread{1.3}\selectfont{} Fractal dimension throughout the lung by Scadding stage. The black solid lines represent the mean estimate for healthy controls; Stage I (solid light blue); Stage II (dashed blue); Stage III (dash-dotted purple); Stage IV (dotted pink).}
  \label{fig:fd1}
\end{figure}

There is also a trend in fractal dimension across healthy controls and stages of sarcoidosis (Figure~\ref{fig:fd1}). In general, healthy subjects have the highest fractal dimension, and sarcoidosis subjects classified as stage IV have the lowest fractal dimension, as we would expect. The trend across stages from highest to lowest FD does not following the staging progression and is as follows: Stage III, Stage I, Stage II, and Stage IV. 

% 
% \subsubsection{Comparison of Radiomic Features}
% Fractal dimension,  Moran's $\mathcal{I}$, and Geary's C all differentiate between subjects with and without sarcoidosis at certain parts in the lung. Figure~\ref{fig:compare1} compares the significance of these biomarkers, adjusted for age, gender, and BMI, with values above 2 indicative of significant regions at a significance level of $\alpha=0.05$. Fractal dimension (black line) has a higher curve than the other biomarkers across all lungs and planes, except for a small region in the sagittal plane. Thus, in general, fractal dimension does better at differentiating across the entire lung; however, in certain regions,  Moran's $\mathcal{I}$ and Geary's C also do well. 

\newpage
\bgroup
\def\arraystretch{1}%
% latex table generated in R 3.3.1 by xtable 1.8-2 package
% Fri Jul 14 10:08:40 2017
\begin{table}[H]
\centering
\caption{Subject Characteristics}
\label{tab:demo}
\begin{tabular}{l|ccc}
  \hline
 & Control & Sarcoidosis & P-value \\ 
  \hline
Sample size &    101\phantom{0} &     58\phantom{0} &  \\ 
  Non-white &      \phantom{00.00}6 (5.9)  &      \phantom{000.00}7 (12.1)  &  \phantom{0}0.230 \\ 
  Male &     \phantom{0000.}32 (31.7)  &     \phantom{00.00}32 (55.2)  &  \phantom{0}0.006 \\ 
  Smoker &   \phantom{000.0}0 (0.0)  &     \phantom{000.0}14 (24.1)  & $<$0.001 \\ 
  Age (yrs) &  \phantom{00}62.11 (9.07) &  \phantom{00}54.05 (9.03) & $<$0.001 \\ 
  Height (cm) & \phantom{0}166.55 (9.19) & \phantom{0}171.89 (9.80) &  \phantom{0}0.001 \\ 
  Weight (kg) &  \phantom{000}77.98 (16.00) &  \phantom{000}85.65 (21.33) &  \phantom{0}0.019 \\ 
  BMI &  \phantom{00}28.02 (4.89) &  \phantom{00}28.84 (6.23) &  \phantom{0}0.389 \\ 
  Heart rate &  \phantom{000}72.10 (11.55) &  \phantom{000}74.55 (12.56) &  \phantom{0}0.225\\ 
   \hline
\end{tabular}
\end{table}
\egroup

\bgroup
\def\arraystretch{1}%
\begin{table}[H]
\centering
\caption{Clinical Characteristics}
\label{tab:clin}
\begin{tabular}{l|ccc}
  \hline
 & Control & Sarcoidosis & P-value \\ 
  \hline
Sample size &    101 &    58 &  \\ 
  \% Pred Post-BD FEV1 & 103.31 (13.81) & 85.22 (19.17) & $<$0.001 \\  
  \% Pred Post-BD FVC &  \phantom{0}99.13 (12.12) & 88.02 (14.60) & $<$0.001 \\ 
   \hline
\end{tabular}
\end{table}
\egroup

\bgroup
\def\arraystretch{1}%
\begin{table}[H]
\centering
\caption[Summary of selected axial slices]{Summary for 70th percentile axial slice of selected persons}
\label{tab:axial}
\begin{tabular}{c|ccc|ccc}
\hline
& & Right Lung & & & Left Lung & \\ 
& Healthy & Sarc A & Sarc B & Healthy & Sarc A & Sarc B\\ 
\hline
D & 2.33 & 2.21 & 2.15 & 2.36 & 2.25 & 2.18 \\ 
$\mathcal{I}$ & 0.89 & 0.91 & 0.95 & 0.87 & 0.89 & 0.90 \\ 
C & 0.09 & 0.07 & 0.04 & 0.10 & 0.08 & 0.05 \\ 
\hline
\end{tabular}\\
\end{table}
\egroup

\bgroup
\def\arraystretch{1}%
\begin{table}[H]
\centering
\caption[Summary of selected coronal slices]{Summary for 40th percentile coronal slice of selected persons}
\label{tab:coronal}
\begin{tabular}{c|ccc|ccc}
\hline
& & Right Lung & & & Left Lung & \\ 
& Healthy & Sarc A & Sarc B & Healthy & Sarc A & Sarc B\\ 
\hline
D & 2.35 & 2.22 & 2.16 & 2.33 & 2.25 & 2.14 \\ 
$\mathcal{I}$ & 0.82 & 0.92 & 0.94 & 0.83 & 0.90 & 0.96 \\ 
C & 0.13 & 0.07 & 0.04 & 0.13 & 0.07 & 0.03 \\ 
\hline
\end{tabular}
\end{table}
\egroup 

\bgroup
\def\arraystretch{1}%
\begin{table}[H]
\centering
\caption[Summary of selected sagittal slices]{Summary for 80th percentile sagittal slice of selected persons}
\label{tab:sagittal}
\begin{tabular}{c|ccc}
\hline
& & Left Lung & \\ 
& Healthy & Sarc A & Sarc B \\ 
\hline
D & 2.37 & 2.25 & 2.14 \\ 
$\mathcal{I}$ & 0.81 & 0.86 & 0.96 \\ 
C & 0.13 & 0.10 & 0.04 \\
\hline
\end{tabular}
\end{table}
\egroup

\newpage

% 
% % Need to update these still. 
% \begin{figure}[H]
% \centering
%   \includegraphics[width=2.25in]{Figures/right_axial_compare.pdf}
%   \includegraphics[width=2.25in]{Figures/left_axial_compare.pdf}
%   \includegraphics[width=2.25in]{Figures/right_coronal_compare.pdf}
%   \includegraphics[width=2.25in]{Figures/left_coronal_compare.pdf}
%   \includegraphics[width=2.25in]{Figures/right_sagittal_compare.pdf}
%   \includegraphics[width=2.25in]{Figures/left_sagittal_compare.pdf}
% \caption[Summary of putative radiomic biomarkers throughout the lungs]{\linespread{1.3}\selectfont{} Effect of disease on putative radiomic biomarkers, adjusted for age, gender, and BMI. Legend: fractal dimension (solid black),  Moran's $\mathcal{I}$ (solid light blue), Geary's C (dashed light blue), and Mat\'{e}rn characteristic length scale (solid purple), smoothing parameter (dashed purple), and partial sill (dotted purple).}
%   \label{fig:compare1}
% \end{figure}

\section{Discussion}
\label{sec:discussion}
In this paper, we demonstrate the ability of spatially-varying, spatial radiomic biomarkers to characterize pulmonary sarcoidosis. The three spatially-varying spatial biomarkers include:  Moran's $\mathcal{I}$, Geary's C, fractal dimension, and Mat\'{e}rn characteristic length scale \& smoothness parameter. Fractal dimension, a measure of self-similarity or image roughness, shows the strongest statistical evidence for differences between sarcoidosis and control images across the whole lung, indicating sarcoidosis CT images may be best characterized by patterns (or rather the lack thereof) of self-similarity and roughness. Interpretation of this finding suggests that when sarcoidosis affects the lung, the inherent self-similarity pattern in the lung is disrupted by disease, resulting in CT findings with less self-similar and more smoothness parameters in sarcoidosis individuals. By calculating the fractal dimension on slices throughout the lung, we are able to objectively and quantitatively show that this disruption (i.e. less self-similarity) is occurring in subjects with sarcoidosis.

While significant differences in fractal dimension between sarcoidosis subjects and healthy controls are seen throughout the entire lung, disease abnormalities are most apparent in the top axial, middle coronal, and outer sagittal regions, corroborating other studies that indicate CT abnormalities are predominately found in the upper and central regions of the lung (Lynch 2003). 

Furthermore, our results show a trend in Scadding stages, with stage III subjects having the highest fractal dimension, in general, followed by stage I, stage II, then stage IV. Since neither CT scan severity nor clinical symptoms necessarily increase with increasing Scadding stages I-III, this is also consistent with our understanding of the Scadding system. Subjects with Scadding stage III differ from the others in that they have no bilateral hilar lymphadenopathy or BHL. From these fractal dimension results, it appears that fractal dimension is sensitive to the larger light areas and patterns in subjects with BHL, and suggests that stage III subjects may visually show lung scans more similar to healthy controls than other Scadding subjects. However, given our small sample sizes across Scadding stages, we were not able to detect a statistical difference.

Moran's  I and Geary's  C, both measures of spatial autocorrelation, also performed well in characterizing sarcoidosis CT findings, particularly in the top axial and outer left/right sagittal regions. With higher  Moran's $\mathcal{I}$ and lower Geary's C, CT slices from sarcoidosis individuals showed significantly more positive spatial autocorrelation than healthy controls. These results are consistent with our hypotheses: when disease worsens, CT nodules begin to conglomerate into larger masses, resulting in more spatial structure. By calculating  Moran's $\mathcal{I}$ and Geary's C on slices throughout the lung, we are able to objectively and quantitatively characterize that this conglomeration (i.e. more positive  spatial  structure) is occurring in subjects with sarcoidosis.

Although similar, Geary's C slightly outperforms  Moran's $\mathcal{I}$, especially in the axial anatomical plane. Since Geary's C is based on weighted squared differences, rather than the absolute value of deviations from the global mean, it is better able to differentiate between smaller scale spatial structure, such as those exhibited by CT nodules. Since more CT nodules become apparent as disease worsens across the whole lung, Geary's C should and does perform slightly better overall. However, when the CT nodules begin to coalesce into larger conglomerate masses or become fibrotic, especially in the top axial region, we should and do observe higher global spatial structure, which  Moran's $\mathcal{I}$ is more suited to measure.  

Our  study  is limited  by  the  number  of CT  scans  on  subjects with  sarcoidosis, especially in Scadding  stage I and III subjects. Since sarcoidosis is a rare disease that has not been frequently studied, it was difficult to obtain these scans. However, we plan to perform these same analyses on the entire population from the GRADS study in the near future. Additionally, there were some demographic discrepancies between the healthy and diseased subjects. However, this was accounted for in the analysis by adjusting for age and gender.  Another limitation in this study was the previous lack of useful ways to implement a functional analysis, given the size of our data.  We  overcame this by using a novel and unbiased binning technique. 

In this study, we have successfully calculated  Moran's $\mathcal{I}$, Geary's C, and fractal dimension on CT slices throughout the lung in the three anatomical planes.  Additionally, we have shown that  Moran's $\mathcal{I}$, Geary's C and fractal  dimension significantly differentiate between subjects with and without sarcoidosis in ways that approximate what we know about the pathology of sarcoidosis. However, we were not able to identify a statistical difference between Scadding stages due to small group sample sizes and lack of power; yet, CT scans from patients with Scadding stages I and III appeared to trend the healthiest from our biomarker analyses, and Scadding stage IV appeared the least healthy. Finally, using functional regression, we were able to identify regions in the lung where CT abnormalities tend to be prevalent. 

Thus, the radiomic measures and techniques presented herein were successfully able to characterize computed tomography images in sarcoidosis. We are optimistic regarding the potential application and use of our methods on other medical imaging modalities and disease datasets in the near future.

Acknowledgements. We are grateful to the GRADS and the COPDGene studies, allowing us to access the data. COPDGene is funded by Award Number R01 HL089897 and Award Number R01 HL089856 from the National Heart, Lung, and Blood Institute. The content is solely the responsibility of the authors and does not necessarily represent the official views of the National Heart, Lung, and Blood Institute or the National Institutes of Health.

\medskip
 %Bibliography

\end{document}